\documentclass[useAMS,usenatbib]{mn2e}
\input{epsf}

\def\lsim{ \lower .75ex \hbox{$\sim$} \llap{\raise .27ex \hbox{$<$}} }
\def\gsim{ \lower .75ex \hbox{$\sim$} \llap{\raise .27ex \hbox{$>$}} }
\def\g{{\tt GALFORM}}

\begin{document}
 
\title
[The structural and photometric properties of early-type galaxies]
{The structural and photometric properties of early-type galaxies in 
hierarchical models} 
 
\author[Almeida  et al.]{
\parbox[t]{\textwidth}{
\vspace{-1.0cm}
C.\,Almeida$^{1}$,
C.\,M.\,Baugh$^{1}$,
C.\,G.\,Lacey.$^{1}$
}
\\
$^{1}$Institute for Computational Cosmology, Department of Physics, 
University of Durham, South Road, Durham, DH1 3LE, UK. 
}
 
\maketitle
\begin{abstract}
We present predictions for the structural and photometric properties of 
early-type galaxies in the cold dark matter cosmology ($\Lambda$CDM) 
from the published semi-analytical galaxy formation models of Baugh et~al. 
and Bower et~al. These calculations were made with the \g~code, which 
tracks the evolution of the disc and bulge components of a galaxy, using 
a self-consistent model to compute the scalelengths. The sizes of galactic 
discs are determined by the conservation of the angular momentum of 
cooling gas. The sizes of merger remnants are computed by applying the 
virial theorem and conserving the binding energy of the progenitors and 
their orbital energy. There are a number of important differences between 
the two galaxy formation models. To suppress the overproduction of 
bright galaxies, the Bower et~al. model employs AGN 
heating to stifle gas cooling, whereas the Baugh et~al. model invokes 
a superwind which ejects cooled gas. Also, in the Baugh et~al. model a 
top-heavy stellar initial mass function is adopted in starbursts.
We compare the model predictions with observational results 
derived from the SDSS. The model enjoys a number of notable successes, 
such as giving reasonable reproductions of the local Faber-Jackson relation 
(velocity dispersion-luminosity), the velocity dispersion-age relation, 
and the fundamental plane relating the luminosity, velocity 
dispersion and effective radius of spheroids. These achievements 
are all the more remarkable when one bears in mind that none of the 
parameters have been adjusted to refine the model predictions. 
We study how the residuals around the fundamental plane relation 
depend on galaxy properties. We examine in detail the physical 
ingredients of the calculation of galaxy sizes in \g, showing which 
components have the most influence over our results. We also study 
the evolution of the scaling relations with redshift. However, in spite 
of the successes, there are some important disagreements between 
the predictions of the model and observations: the brightest model 
spheroids have effective radii smaller than observed and the zero-point 
of the fundamental plane shows little or no evolution with 
redshift in the model.
\vspace{0.75cm}
\end{abstract}

\section{Introduction}

Remarkably tight correlations exist between the structural and 
photometric properties of galaxies. Across the Hubble 
sequence there is a strong dependence of luminosity on either the 
rotation speed of galactic discs \citep{tully-fisher} or the 
velocity dispersion of spheroids \citep{faber}. Other scaling 
relations observed for early-type galaxies include those between 
colour and magnitude \citep{sandage.a,sandage.b}, colour and velocity 
dispersion \citep{b05}, radius and luminosity \citep{sandage}, and radius and 
surface brightness \citep{kormendy}. Some of these correlations can be 
combined into a ``fundamental plane'' which connects the effective 
radii, velocity dispersions and luminosities of ellipticals 
\citep{djor,dress,b03I}.  

The existence of these scaling relations and their tightness encode 
clues about the formation and evolution of elliptical galaxies.  
For example, the existence of a fundamental plane can be understood 
by applying the virial theorem to a gravitationally bound stellar spheroid 
in dynamical equilibrium, after making the assumption that ellipticals of 
different sizes have the same structure (homology) and a constant 
mass-to-light ratio. The deviation of the observed fundamental plane 
from this prediction can therefore be driven by variations in the 
mass-to-light ratio across the early-type population or by a non-uniformity 
of the structure of ellipticals, referred to as structural non-homology, or a 
combination of these two effects \citep{ciotti, bertin, trujillo}. 
At first sight, the small scatter around the observed correlations 
would appear to pose a challenge to hierarchical 
galaxy formation models, since the variety of merger histories in the models 
would lead one to expect to a corresponding scatter in the properties of 
early-type galaxies.

In this paper, we use a semi-analytical approach to model the 
properties of elliptical galaxies. Such models predict the star 
formation and merger histories of galaxies \citep[for a review 
of this class of model, see][]{baugh06}. 
In general, two channels are considered 
for the formation of spheroids: galaxy mergers or secular evolution of 
the disc. We will describe the formation of discs and bulges in more 
detail in Section 2. The first attempts to track the disc and bulge 
of a galaxy separately simply recorded the mass and luminosity in each 
component (Baugh, Cole \& Frenk 1996; Kauffmann 1996). 
The models have now progressed to a state where 
detailed predictions can be produced for the structural properties of 
galaxies in addition to their stellar populations. 
\citet{cole00} introduced a model for the sizes of the disc and 
spheroid components of galaxies: the size of a galactic disc is calculated 
by assuming conservation of the angular momentum of the gas 
as it cools and collapses in the halo; the size of the spheroid is derived 
by applying conservation of binding and orbital energy, and by applying 
the virial theorem to the merging galaxies. 
The Cole et~al. scheme also takes into account the 
gravitational force of the dark matter and the reaction of the dark 
matter halo to the presence of the baryons (see Section 2 for further 
details).

Cole et~al. tested their model for the sizes of galactic discs against 
the observed distribution of disc scale lengths estimated by \citet{jong}, 
and verified that the predictions of their fiducial model  were in excellent 
agreement with the observations. 

Cole et~al. did not test their prescription for predicting the size of 
galactic spheroids. This is the focus of our paper. In related studies, 
Gonzalez et~al. (2007, in prep.) test the predictions of galaxy scale 
lengths for disks and bulges at $z=0$ and Coenda et~al. (2007, in prep.), 
look at the evolution of galaxy sizes.

\citet{h03} used a similar scheme to that outlined by Cole et~al. 
to compute the sizes of spheroids in the {\tt GALICS} model. 
However, these authors adopted a less realistic model for the scale size 
of galactic discs. In common with many semi-analytical models, they 
assumed that the scale size of a disc is related to the virial radius of 
the host dark matter halo by $r_{\rm D} = \lambda R_{200}/2$, where $\lambda$ 
is the dimensionless spin parameter for the dark matter halo, which 
quantifies its angular momentum, and $R_{200}$ is the halo virial radius.
This ignores the self-gravity of the baryons and the contraction they 
produce in the central regions of the dark matter halo. 
Several papers have considered the origin of the fundamental plane and the 
role of gas-rich and gas-poor mergers using numerical simulations, which 
follow the dark matter and baryons \citep{kobayashi, dekel, rob, boylan}.

In Section 2, we summarize our model, explaining the ingredients which 
are particularly pertinent to the formation of galactic spheroids.  
We first compare our predictions to the sample of early-type galaxies drawn 
from the Sloan Digital Sky Survey by \citet{b05}: the selection criteria are 
described in Section 3 and the comparisons between our model predictions and 
the data are given in Section 4. In Section 5, we explore the sensitivity of 
our model predictions to various physical ingredients of the models. The evolution 
with redshift of the model predictions for the scaling relations is presented 
in Section 6. Our conclusions are given in Section 7. 

\section{The galaxy formation model}
\label{section:model}

A comprehensive overview of the {\tt GALFORM} model of galaxy 
formation and the philosophy behind semi-analytical modelling can be found 
in \citet{cole00} \citep[see also the review by ][]{baugh06}. Important extensions to the model are 
described in \citet{benson02} 
and \citet{b03}. In this paper, we focus on the predictions of the 
model introduced by \citet{baugh05}. These authors put forward the first 
fully consistent hierarchical galaxy formation model which was able to explain 
the observed number counts of sub-mm sources and the luminosity function of 
Lyman-break galaxies, at the same time as reproducing the observed properties 
of the low redshift galaxy population. In some instances, we also show 
predictions from the model described by \citet{bower}, which includes 
feedback processes associated with the 
accretion of material onto a supermassive black hole, using 
the model of black hole growth explained in \citet{rowena}. 
Predictions from the Bower et~al. model 
can be downloaded over the internet \citep[see][]{lemson}.

We now give a brief overview of the {\tt GALFORM} model, referring the 
reader to the references given in the previous paragraph for further details. 
We then outline some of the differences between the Baugh et~al. and Bower et~al. 
models. Finally, we recap some of the ingredients of the model which are 
particularly important for determining the masses and sizes of galactic 
spheroids and discs.
 
The aim of the {\tt GALFORM} model is to make an {\it ab initio} calculation 
of the formation and evolution of the galaxy population, set in the context 
of a cosmological model in which structures in the dark matter form 
hierarchically through gravitational instability. 
The main physical processes which we incorporate into the model 
are the following: 
(i) The hierarchical merging and collapse of dark matter haloes. 
(ii) The radiative cooling of shock heated gas. 
(iii) Quiescent star formation in discs.  
(iv) Feedback processes driven by supernovae and by the accretion of 
material onto supermassive black holes in the case of the model of 
\citet{bower}. 
(v) The effect of a photoionizing background of radiation on the 
intergalactic medium and on galaxy formation\citep[see][]{benson02}.
(vi) The chemical enrichment of the gas and stars. 
(vii) The decay of the orbits of galactic satellites due to dynamical friction.
This can lead to mergers between galaxies which can trigger bursts 
of star formation and a change in galaxy morphology (see the next subsection). 
The model generates a star formation history and a galaxy merger history for a 
representative population of galaxies at any epoch. Each galaxy is split into two 
components, a disc and a bulge. The formation of these components is discussed 
in the next subsection.

For completeness, we now give a list of the primary differences between 
the \citet{baugh05} and \citet{bower} models, even though the 
Baugh et~al. model is the main focus of the comparisons presented 
in the paper:
\begin{enumerate}
\item {\it Dark matter halo merger trees.} The Bower et~al. model 
utilizes merger histories for dark matter haloes drawn from the 
Millennium Simulation of the hierarchical clustering of dark matter 
in a $\Lambda$CDM universe \citep{springel}. The  
simulation covers a volume of $0.125 h^{-3} {\rm Gpc}^{3}$. 
The mass resolution of the trees extracted from the simulation 
is $1.72 \times 10^{10}h^{-1} M_{\odot}$. The Baugh et~al. 
model is not set in the context of an N-body simulation. Instead, 
a representative sample of galaxies is constructed by considering 
a grid of dark halo masses. For each mass on the grid, realizations of 
merger trees are generated using a Monte Carlo algorithm based on 
extended Press-Schechter theory \citep[see][]{cole00}. 
The mass resolution used in the Monte Carlo trees is a factor of three 
better than that of the trees drawn from the Millennium. \citet{helly} 
compared model predictions obtained using Monte Carlo or N-body merger 
trees and reached the conclusion that the results are very similar for bright 
galaxies, with the two prescriptions giving divergent answers for the 
luminosity function at faint luminosities; for the resolution of 
the Millennium, the predictions for the luminosity function are robust 
down to around three magnitudes fainter than the break in the 
luminosity function.
\item {\it The suppression of bright galaxy formation.} Hierarchical 
models tend to produce too many bright galaxies unless some physical 
mechanism is invoked to regulate the formation of massive galaxies. 
The Baugh et~al. and Bower et~al. models do this 
in different ways. Baugh et~al. adopt a superwind driven by star formation 
\citep[see][]{b03}. In this case, the wind drives cold gas out of the 
galactic disc and out of the gravitational potential well of the dark halo. 
The effectiveness of the wind depends upon the depth of the potential well. 
Such winds have been observed in massive galaxies, with inferred mass ejection 
rates which are comparable to the star formation rate 
\citep[e.g.][]{pettini, wilman}. In the Bower et~al. model, the cooling of gas 
in quasi-static hot gas haloes is suppressed, effectively cutting off the 
``fuel supply'' for star formation. These are haloes in which the cooling 
time of the gas exceeds the free fall time within the halo. The cooling is 
quenched by the energy injected into the hot halo by the accretion of 
mass onto the central supermassive black hole in the galaxy. The growth 
of the black hole is followed using the model described by \citet{rowena}. 
\item {\it Quiescent star formation in discs.} The scaling with redshift 
of the timescale for quiescent star formation in galactic discs is different 
in the two models. Both models allow the star formation timescale to depend 
upon some power of the circular velocity of the disc, and multiply this by 
an efficiency factor (eqn. 4.14 of Cole et~al.). Baugh et~al. assume 
that the efficiency factor is independent of redshift, whereas in the 
Bower et~al. model this factor scales with the dynamical time of the galaxy. 
Dynamical times are shorter at high redshifts, so the quiescent star formation 
timescales are shorter in the Bower et~al. model at high redshift than 
they are in the Baugh et~al. model. 
\item {\it Star formation in bursts.} Globally, the two models display 
somewhat different star formation densities in bursts. Furthermore  
bursts are triggered in different ways in the models. Baugh et~al. only consider bursts 
resulting from galaxy mergers. A burst may accompany a major merger in which 
a galaxy accretes a satellite of a comparable mass to its own, or a minor 
merger in which a gas rich disc is hit by a much smaller satellite. 
Bower et~al. consider this mode of initiating star bursts, but also 
incorporate bursts which occur when discs become dynamically unstable. 
The need for this additional channel for bursts is driven by the need to 
build up black hole mass, so that cooling flows can be suppressed in massive 
haloes. One final difference to note between star 
bursts in the two models is that Baugh et~al. invoke a flat initial 
mass function (IMF) for stars produced in a burst, whereas Bower et~al. 
adopt a standard IMF (Bower et~al. adopt a Kennicutt (1998) IMF in 
all modes of star formation; Baugh et~al. use the Kennicutt IMF in quiescent 
star formation).
 
\item{\it Background cosmology.} Baugh et~al. adopt the parameters of the 
concordance $\Lambda$CDM model: matter density, $\Omega_{0}=0.3$, 
cosmological constant, $\Lambda_{0} = 0.7$, baryon density, 
$\Omega_{b}=0.04$ and a normalization of density fluctuations given 
by $\sigma_{8}=0.93$. Bower et~al. use the cosmological parameters of 
the Millennium simulation, which are in better agreement with the 
constraints from the anisotropies in the cosmic mircowave background 
and large scale galaxy clustering \citep[e.g.][]{ariel}: 
$\Omega_{0}=0.25$, $\Lambda_{0} = 0.75$, $\Omega_{b}=0.045$ and 
$\sigma_{8}=0.9$. The power spectrum of density fluctuation used in 
the Millennium has somewhat more large scale power and less small 
scale power than that in the concordance $\Lambda$CDM model.
\end{enumerate}

One difference between the predictions of the two models is the amount 
of star formation which takes place in starbursts. Baugh et~al. calculate 
that 30\% of all star formation in their model takes place in starbursts 
driven by galaxy mergers. However, due to the high recycled fraction in 
starbursts as a consequence of a top-heavy IMF, only 7\% of
this mass is locked up in long lived stars, and yet bulges account for
around 50\% of the global stellar mass in this model. The re-assembly of 
stars which were produced with the standard IMF used in quiescent star 
formation is therefore the primary source of stellar mass in bulges 
(Baugh, Cole \& Frenk 1996; de Lucia et~al. 2006; de Lucia \& Blaizot 2006). 
In the Bower et~al. model, the amount of star formation triggered by galaxy 
mergers is lower than in the Baugh et~al. model for a combination of 
reasons: 
1) Bower et~al. use a quiescent star formation timescale which depends 
on the dynamical time, and is thus shorter at high redshift. Galactic 
disks at high redshift are therefore gas poor in the Bower et~al. model 
compared with those in the Baugh etal model, so there is less fuel for 
the starbursts. 
2) Minor mergers do not trigger bursts of star formation in the Bower et~al. 
model, as is the case in the Baugh et~al. model. 
In the Bower et~al. model, disk instabilities account for over half of 
the star formation in bursts. Due to the choice of a standard IMF in all 
modes of star formation in this model, bursts are responsible for 
producing around a quarter of the mass in bulge. Again, the re-assembly of 
stellar mass made in galactic disks is the main source of spheroid stars.

\subsection{The formation of spheroids and discs}

We assume that gas initially settles into a rotationally supported disc when it cools from 
the hot halo. This gas eventually turns into stars in the quiescent star formation mode. 
The effective timescale on which the star formation takes place does not depend upon the 
dynamical time of the disc in the model of \citet{baugh05}, 
but does have some dependence on the circular velocity of the disc.  

The formation of galactic spheroids takes place through two channels: 
galaxy mergers and the instability of galactic discs. Baugh et~al. 
only consider the galaxy merger mode of spheroid formation; Bower 
et~al. consider both mechanisms. 

The consequences of a galaxy merger are characterized by the ratio, 
$R = m_{\rm sat}/m_{\rm central}$, of the mass of the accreted satellite 
galaxy ($m_{\rm sat}$) to the mass of the primary or central galaxy in 
the halo ($m_{\rm central}$), onto which the satellite is accreted. 
The mass ratio $R$ is compared to two thresholds, $f_{\rm ellip}$ 
and $f_{\rm burst}$, to establish the  severity of the merger \citep[see, for example,][]{bournaud}. 
These thresholds are model parameters. If $R>f_{\rm ellip}=0.3$, then the 
galaxy merger is termed ``major''. 
In the case of major mergers, the disc of the primary galaxy is destroyed. All stars are 
transferred to the spheroid component and any cold gas present participates in 
a burst of star formation which adds stars to the spheroid. 
If the mass ratio of satellite to primary falls between the thresholds, 
i.e. $f_{\rm burst}<R<f_{\rm ellip}$, then the stellar disc of the 
primary survives and the stars from the accreted satellite are added 
to the spheroid. In this case, if the primary is also gas rich, 
that is if cold gas accounts for at least 75\% of the total mass in the disc, 
then we assume that the accretion of the satellite induces an instability which 
drains the primary disc of cold gas, leading to a burst of star formation in the spheroid. 

In our model, a galaxy can move in either direction along the Hubble 
sequence \citep{baugh96}. The accretion of gas from the hot halo 
and subsequent quiescent star formation leads to a late-type (disc dominated) 
galaxy. A major merger between two such galaxies produces a descendent 
galaxy which jumps to the opposite end of the Hubble sequence, becoming 
an early-type galaxy (bulge dominated). Further accretion of cooling gas 
allows the galaxy to grow a new disc around its bulge, moving the galaxy 
back towards the late-type part of the sequence. 

In addition to the merger mode of spheroid production, Bower et~al. 
also consider the secular production of bulges from discs which are 
unstable due to their strong self-gravity. This mode is most important 
in less massive galaxies. 

\subsection{The scale lengths of the disc and bulge components of galaxies}
\label{ssec:sizes}

We assume that discs have an exponential profile, with a half-mass radius 
given by $r_{\rm disc}$, and bulges have a $r^{1/4}$ profile in projection, 
with a half mass in 3D given by $r_{\rm bulge}$. 

The scalelength of the disc is determined by the angular momentum of 
the halo gas, which arises due to the tidal torques which act during the 
formation of the halo. The angular momentum of the halo gas is quantified 
by the dimensionless spin parameter, $\lambda$; this quantity is assumed 
to follow a log-normal distribution matching the results of N-body simulations 
\citep[see][for details]{cole00}. We assume that the angular momentum of the gas 
is conserved as it cools to form a rotationally supported disc 
\citep[see][for a discussion of this assumption]{okamoto}.

Spheroids are formed in galaxy mergers or through disc instabilities 
as outlined in the previous subsection. The size of the spheroid resulting 
from a galaxy merger, $r_{\rm m}$, is determined by applying the 
conservation of energy and the virial theorem \citep[see Section 4.4.2 of][]{cole00}: 
\begin{equation}
\label{eq1}
\frac{(M_1+M_2)^2}{r_{\rm m}} = \frac{M_1^2}{r_1} + \frac{M_2^2}{r_2} + 
\frac{f_{\rm orbit}}{c}\frac{M_1M_2}{r_1+r_2}, 
\end{equation}
where $M_i$ represents the total mass (stellar, cold gas and dark matter) 
of one of the merging objects, within $r_{\rm i}$ , and the form factor $c$ 
and the parameter $f_{\rm orbit}$ are related to the self-binding energy and 
orbital energy by
\begin{eqnarray}
E_{\rm bind} & = & -c \frac{GM_i^2}{r_i} \\
E_{\rm orbit} & = & -\frac{f_{\rm orbit}}{2}\frac{GM_1M_2}{r_1+r_2}.
\end{eqnarray}
For simplicity, we adopt $c=0.5$ and $f_{\rm orbit} = 1$. Later on, 
we explore the impact on our predictions of varying $f_{\rm orbit}$.
Similar arguments are applied to calculate the scale size of the spheroid 
which results from an unstable disc \citep[see Section 4.4.3 of][]{cole00}.

Once the scale lengths of the disc and bulge components have been 
calculated as outlined above, we next take into account the selfgravity 
of the baryons and the contraction of the dark matter halo in response 
to the gravity of the condensed baryons. New radii are computed for the 
disc and bulge by applying an adiabatic contraction of the disc, bulge 
and dark matter components \citep{blumenthal, jesseit}. In the case of the 
disc, the total specific angular momentum is conserved. The bulge and dark 
matter halo are not rotationally supported. Nevertheless, it is useful  
to define an equivalent circular velocity using the velocity dispersion of 
each of these components, and, using this, to define a quantity which we refer to 
as a pseudo-angular momentum. For the bulge the pseudo-angular momentum is 
given by: $j_{\rm bulge} = r_{\rm bulge}V_c(r_{\rm bulge})$. This quantity 
is conserved during the adiabatic contraction. A similar quantity is conserved 
for the dark matter. 

In the case of the secular growth of spheroids, we again apply the conditions 
of virial equilibrium 
and energy conservation as in Eq.~\ref{eq1}, defining the component 1 as 
the galactic bulge, $M_1= M_{\rm bulge}$, $r_1= r_{\rm bulge}$, and 
the component 2 as the unstable disc, $M_2= M_{\rm disc}$, $r_2= r_{\rm disc}$. 
After the calculation of the radius of the new bulge, we readjust adiabatically the 
spheroid and halo terms to reach the new equilibrium.

\section{The selection of an early-type sample}

We first compare our model predictions against the scaling relations and statistics  
of the sample of early-type galaxies constructed from the SDSS by 
\citet{b03I,b05}. These authors measured relations between luminosity and 
various properties of early-type galaxies such as velocity dispersion, 
effective radius, effective mass, effective density and surface 
brightness \citep{b03II,b05}. The sample was also 
used to measure the luminosity function of early-type galaxies \citep{b03II}, 
the fundamental plane \citep{b03III} and the colour-magnitude/colour-velocity 
dispersion relations \citep{b03IV,b05}. 

For a complete description of the construction of the early-type sample 
from the SDSS, we refer the reader to the above  papers. Below we give a 
summary of the selection criteria applied by \cite{b05} (hereafter Ber05). Galaxies 
are included in the sample if they have:  

\begin{enumerate}
\item Redshift $z \leq 0.3$, with a median $z_{\rm med} = 0.13$.
\item Apparent r-band Petrosian magnitudes in the range 
      $14.5 < r_{\rm petro} < 17.75$.
\item {\tt eclass} $< 0$. The {\tt eclass} value is a classification of the 
      spectral type of a galaxy derived from a principal component 
      decomposition of its spectrum \citep{connolly}. Ber05 chose  
      negative values of {\tt eclass} as this corresponds to spectra in 
      which absorption lines dominate, characteristic of 
      early-type galaxies. 
\item {\tt fracDev}$ > 0.8$, computed using the $r-$band image. The value 
      of {\tt fracDev} is an indicator of morphology. It is calculated in two steps, in turn. 
      First, the best fit de Vaucouleurs and exponential profiles to the galaxy image 
are found. Second, using the scale lengths of the best fit profiles found in step one, the best 
fit linear combination of the disc and bulge profiles is derived. The contribution of the 
de Vaucouleurs profile to this linear combination is the value of {\tt fracDev}.
\end{enumerate}

We attempt to reproduce these selection criteria by 
imposing the following conditions on {\tt GALFORM} galaxies:
\begin{enumerate}
\item We generate a population of galaxies at an output redshift of 
$ z = 0.13$, the median redshift of the Ber05 sample. In our comparisons, 
we consider only Ber05 galaxies that lie within the redshift interval 
$0.11 < z \leq 0.15$, close to the median redshift. This additional 
selection in redshift reduces the size of the observational sample 
by a factor of $\sim 4$ to $\sim 11000$ objects. 
\item We use total magnitudes to select a sample of model galaxies. 
We apply the same apparent magnitude limits which are used for the data. 
This is a reasonable approach as the difference between total and 
Petrosian magnitudes is typically smaller than 0.2 mag \citep{graham}.
We have also computed Petrosian magnitudes for our model galaxies and  
find that using the Petrosian magnitudes in place of total magnitudes does not make 
a significant difference to our results. 
\item At present, {\tt GALFORM} does not produce spectra with absorption 
line features. Therefore we cannot directly calculate a value for the spectral 
parameter {\tt eclass}. Instead, we use the $g-r$ colour which is more readily 
predicted for model galaxies. In Fig.~\ref{fig:eclass}, we use the 
SDSS DR4 to show that there is a good correlation between $g-r$ colour 
and {\tt eclass}. We retain galaxies with $g-r \ga 0.8$; Fig.~\ref{fig:eclass} 
shows that more than 95\% of the galaxies with a negative value 
for {\tt eclass} are selected by this colour cut.  
\begin{figure}
{\epsfxsize=8.5truecm
\epsfbox[18 144 592 718]{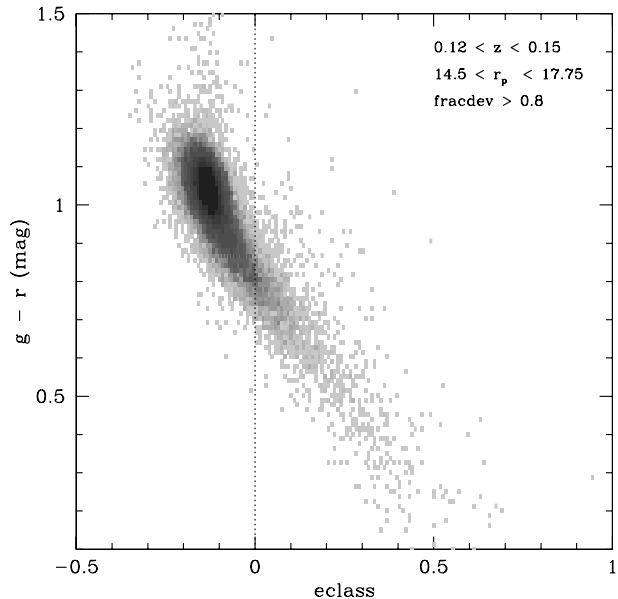}}
\caption{
The relation between $g-r$ colour and {\tt eclass}, for a sample of galaxies 
selected from SDSS DR4 in the redshift range $0.12 < z < 0.15$ 
and with Petrosian magnitudes in the interval 
$14.5 < r_{\rm petro} < 17.75$. The dotted line represents the selection 
applied to the observational data, {\tt eclass} $< 0$, by \citet{b05}: 
more than 95\% of these galaxies have $g-r$ $>0.8$.}
\label{fig:eclass}
\end{figure}
 \item We compute the value of {\tt fracdev} in the a similar way as was done for 
the SDSS galaxies. We assume that the model bulges follow a de 
Vaucouleurs profile and the discs an exponential profile; these profiles 
describe the distribution of stellar mass, and so are independent of the 
passband. We adopt a cut on {\tt fracdev} $>0.8$. In 
Section~4.1, we explore the impact on our predictions of replacing the 
cut in the value of {\tt fracdev} with a simple cut on the bulge-to-total 
luminosity ratio of the model galaxies. 
\item Due to limitations of the SDSS data we also set a surface brightness 
      threshold, $\mu_e < 24.5$ mag arcsec$^{-2}$.
\end{enumerate}

\section{Results}

In this section, we compare the predictions of the \citet{baugh05} 
model with observational data for early-type galaxies derived from 
the SDSS sample of Ber05.

\subsection{The luminosity function of early-type galaxies}
The luminosity function is perhaps the most fundamental statistical 
description of the galaxy population. Later on, we perform fits to 
the fundamental plane of early-type galaxies in our model. The results 
for this fit are sensitive to the abundance of galaxies as a function 
of luminosity, so it is imperative that the model reproduces 
the observed luminosity function closely. 

\begin{figure}
{\epsfxsize=8.5truecm
\epsfbox[18 144 592 718]{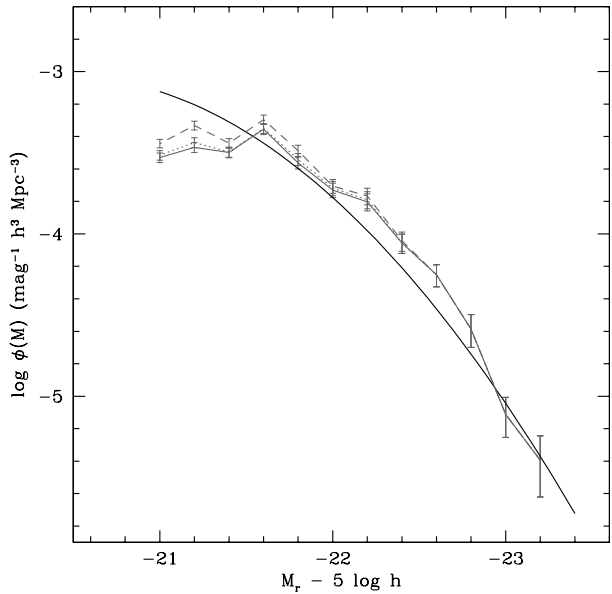}}
\caption{
The luminosity function of early-type galaxies at $z = 0.13$. 
The solid black line shows a fit to the luminosity function of the SDSS 
sample of \citet{b05}.
The results for the {\tt GALFORM} sample are plotted using grey lines. 
The solid grey line shows our standard early-type galaxy selection, as outlined 
in Section 3. The errorbars show Poisson errors due to the finite number 
of galaxies simulated.  
The other lines show how the luminosity function varies when, instead of 
using {\tt fracdev} $>0.8$, the bulge-to-total $r$-band luminosity ratio is used; 
the dashed line shows the results for B/T$_r > 0.5$ and the dotted line 
for B/T$_r > 0.8$. 
}
\label{fig:lf}
\end{figure}

In Fig.~\ref{fig:lf}, we compare the predictions of the {\tt GALFORM} model 
for the $r$-band luminosity function of early-type galaxies with the estimate 
from the SDSS sample of Ber05. The luminosity function of SDSS early-types is  
well described by a Gaussian form: 
$\phi(M)\, dM = \phi_{\star}/\sqrt{2 \pi \sigma^2} 
\exp\{-(M - M_{\star}+Qz_i)^2/(2 \sigma^2)\}$, 
where $(\phi_{\star}, M_{\star}, \sigma, Q) = 
(1.99\times 10^{-3}{\rm Mpc}^{-3}, -21.15, 0.841,0.85)$ 
respectively (note, Ber05 assume $h=0.7$).
The model predictions are in reasonably good agreement with the 
luminosity function estimated from the data. 

We also show, in Fig.~\ref{fig:lf}, the effect of changing the criteria used to 
select early-type galaxies in the model. In our standard 
selection, the primary indicator of morphology is {\tt fracdev}  (see Section 3). 
We have also explored using the bulge-to-total 
luminosity ratio in the $r$-band, B/T$_r$, in place of {\tt fracdev} \citep{baugh96}. 
The results for cuts of B/T$_r>0.8$  and B/T$_r>0.5$ are shown by 
the dotted and dashed lines respectively in Fig.~\ref{fig:lf}. The luminosity 
function derived using B/T$_r>0.8$ is remarkably similar to the one obtained  
using {\tt fracdev} $ >0.8$ (shown by the solid grey line). 

In our subsequent comparisons with the SDSS sample of early-types, we 
assign each model galaxy a weight which depends upon luminosity, such 
that the luminosity function of early-types in the model is forced to 
reproduce exactly the luminosity function of the data. 

 \subsection{The Faber-Jackson and $\sigma$-age relations}
The Faber-Jackson (hereafter FJ) relation was one of the first 
scaling relations to be discovered for early-type galaxies \citep{faber}. 
This relation indicates that luminosity is a strong function of 
velocity dispersion, $\sigma$, with brighter early-types displaying 
larger velocity dispersions. Observational studies suggest that this relation 
is approximately given by $L \propto \sigma^4$: \citet{forbes}, using a 
local sample of early-type galaxies, found 
$L_B \propto \sigma^{3.9}$ in the B-band, while \citet{pahre} reported 
$L_K \propto \sigma^{4.1}$ in the K-band. 
These results also indicate that the FJ relation is essentially 
independent of wavelength. 
In the case of SDSS early-type galaxies, \citet{b03II} confirmed these 
earlier results, finding  
$L_r \propto \sigma^{3.91}$ in the r-band at $z = 0$, with no 
significant differences in slope apparent in the g, i or z-bands. 

\begin{figure}
{\epsfxsize=8.5truecm
\epsfbox[18 144 592 718]{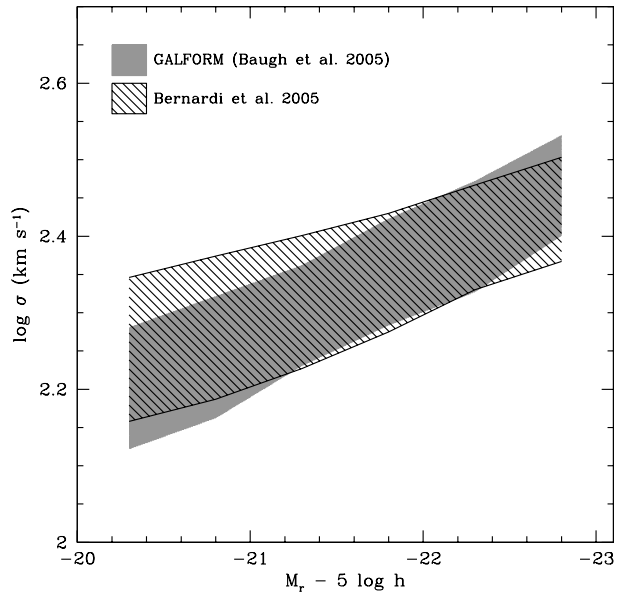}}
\caption{The Faber-Jackson relation between luminosity and velocity 
dispersion. The \g~prediction is shown in grey and the hatched 
shaded region shows the relation for the Ber05 sample. 
The shaded and hatched regions connect the 10 and 90 percentiles. 
The one-dimensional velocity dispersion was calculated using 
$\sigma_{\rm 1D} = (1.1/\sqrt{3}) V_{\rm c, bulge}$, as explained in 
the text.
}
\label{fig:sigma}
\end{figure}

Fig.~\ref{fig:sigma} shows the velocity dispersion-magnitude 
relation predicted by \g. The one-dimensional velocity dispersion 
is calculated using $\sigma_{\rm 1D} = (1.1/\sqrt{3}) V_{\rm c, bulge}$, 
where $V_{\rm c, bulge}$ is the effective circular velocity of the 
bulge, which is assumed to be isotropic \citep[]{frenk, cole94}. The 
factor of 1.1 is an empirical correction which Cole et~al. employed to 
map data for elliptical galaxies onto the Tully-Fisher relation for 
spiral galaxies. Fig.~\ref{fig:sigma} shows that retaining this factor 
gives good agreement with the observed FJ relation.  
We find no change in the predictions if we consider, instead, the 
effective circular velocity of the disc and bulge combined. 
This is a consequence of our 
selection which ensures that the model galaxies we consider are bulge 
dominated, as shown by Fig.~\ref{fig:lf}. 
We find reasonably good agreement between model predictions 
and the FJ relation observed for the Ber05 sample, albeit with a 
shallower slope,  $L_r \propto \sigma^{3.2\pm 0.1}$ (note, we 
plot $\sigma$ on the y-axis). Whilst the slope of the predicted FJ 
relation is formally at odds with that measured by Ber05, it is clear from 
Fig.~3 that the velocity dispersion of the model galaxies would change 
by relatively little even in the case of perfect agreement between the 
observed and predicted slopes, given the limited range of magnitudes 
plotted. We find little dependence of the slope of the FJ relation 
on passband, in agreement with observations.  

\begin{figure}
{\epsfxsize=8.5truecm
\epsfbox[18 144 592 718]{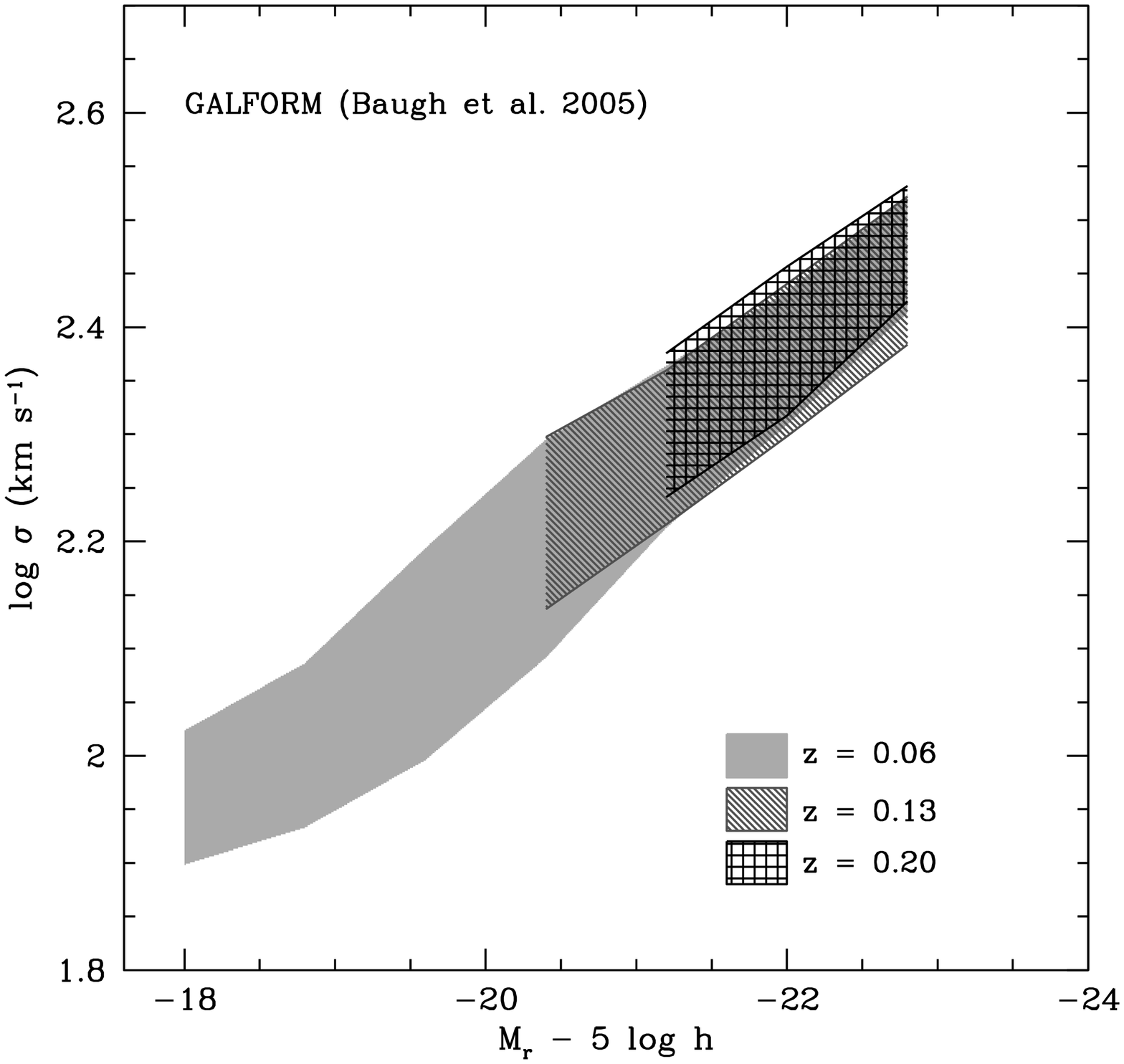}}
{\epsfxsize=8.5truecm
\epsfbox[18 144 592 718]{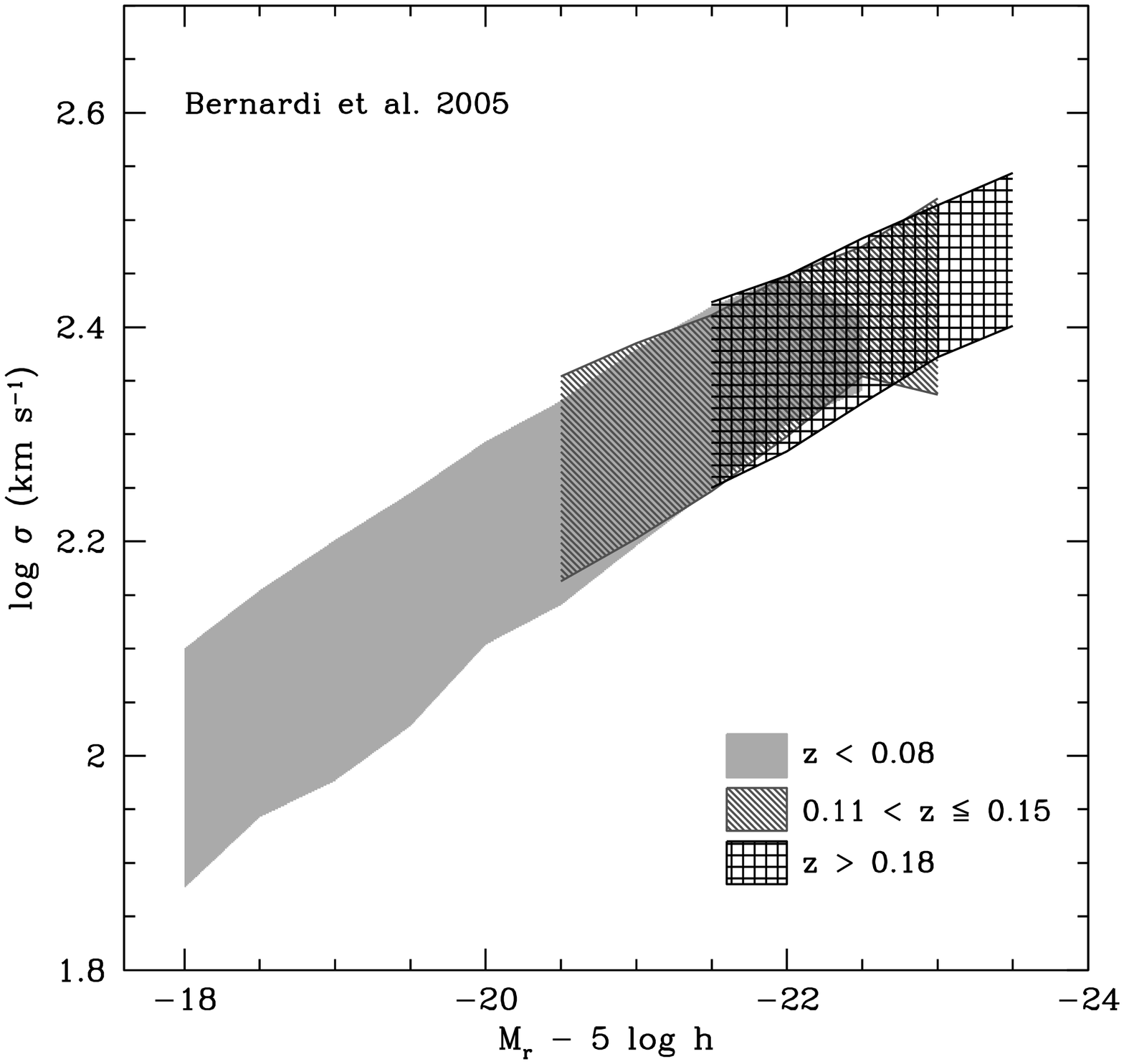}}
\caption{
The evolution of the Faber-Jackson relation for early-type galaxies. 
The upper panel shows the model predictions at redshifts $z=0.06$, $z=0.13$ 
and $z=0.20$, which are the median redshifts of the observational subsamples 
shown in the lower panel. Shaded regions connect the 10 and 90 
percentiles of the distributions.
}
\label{fig:sigma.evo}
\end{figure}

The evolution of the FJ relation with redshift is plotted in 
Fig.~\ref{fig:sigma.evo}. Here, we have chosen output redshifts 
in \g~to match the median redshifts of the Ber05 redshift subsamples: 
for SDSS galaxies with $z < 0.08 $ we use $z_{\rm med} = 0.06$ for the 
model galaxies, and for SDSS galaxies with $z > 0.18 $ we 
set $z_{\rm med}  = 0.20$. Little evolution is observed in the 
zero-point of the FJ relation with redshift, a trend which is 
reproduced by the model predictions. The shift to brighter magnitudes 
with increasing redshift is simply a reflection of the fixed apparent 
magnitude limit of the SDSS. As we shall see in next section, this 
absence of evolution is actually expected at these redshifts due to 
the cancellation of two different evolutionary effects. 

\begin{figure}
{\epsfxsize=8.5truecm
\epsfbox[18 144 592 718]{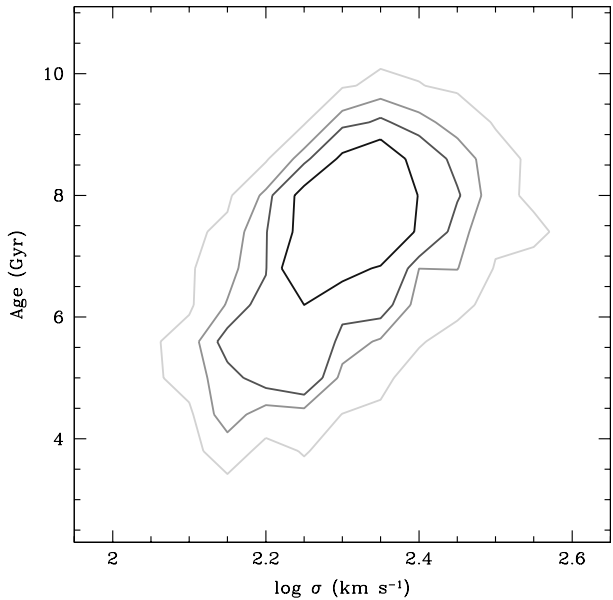}}
\caption{
The $r$-band luminosity-weighted age of the stellar population 
as a function of the velocity dispersion predicted by \g. 
The model galaxies are at redshift $z=0.13$ and are selected 
in a similar way to the observational sample of Ber05. 
The contours enclose $35\%$, $72\%$, $89\%$ and $99\%$ of 
the number density respectively, moving from black to grey.
}
\label{fig:dsigma.age}
\end{figure}

Finally, in Fig.~\ref{fig:dsigma.age}, we plot the luminosity-weighted age 
of the stellar population, computed in the $r$-band,  as a function of the 
velocity dispersion. Some authors have argued that a correlation 
exists between these quantities, which has implications for the 
scatter in the FJ relation \citep{forbes, nelan}.
Fig.~\ref{fig:dsigma.age} reveals that velocity dispersion 
and luminosity-weighted age are indeed correlated in the model, 
with galaxies which have larger velocity dispersions also 
displaying older stellar populations. A linear fit to the model 
predictions shows that 
$\mbox{Age} \propto \sigma^{0.58\pm 0.02}$, which is in excellent 
agreement with recent determination by \citet{nelan}, who found  
$\mbox{Age} \propto \sigma^{0.59\pm 0.13}$. Furthermore, 
we verify that the inclusion of AGN feedback, as implemented 
by \citet{bower}, does not change this relation substantially, 
giving $\mbox{Age} \propto \sigma^{0.51\pm 0.03}$. 
At first sight, these predictions seem to contradict those presented, 
for a different observational selection, by Nagashima et~al. (2005b). 
However, it is important to note that, at least in the case of the 
model galaxies, the slope and scatter of the $\mbox{Age}-\sigma$ 
relation are very sensitive to the selection criteria applied. 

 \subsection{Radius-Luminosity Relation}
Another component of the fundamental plane of early-type galaxies is 
the relation between radius and luminosity; galaxies with larger radii 
are more luminous. This was originally of interest for use in distance 
scale measurements and cosmological tests \citep{sandage}. 
Different studies indicate that this relation varies slightly 
with wavelength. For example, \citet{shade} determined 
$L_B \propto r_e^{1.33}$ in the B-band and \citet{pahre} found 
$L_K \propto r_e^{1.76}$ in the K-band. For SDSS early-type galaxies, 
\citet{b03II} reported $L_g \propto r_e^{1.50}$ in the g-band and 
$L \propto r_e^{1.58}$ in the r, i and z-bands, which is consistent 
with the variation of the slope of this relation with wavelength 
suggested by the earlier determinations. 

\begin{figure}
{\epsfxsize=8.5truecm
\epsfbox[18 144 592 718]{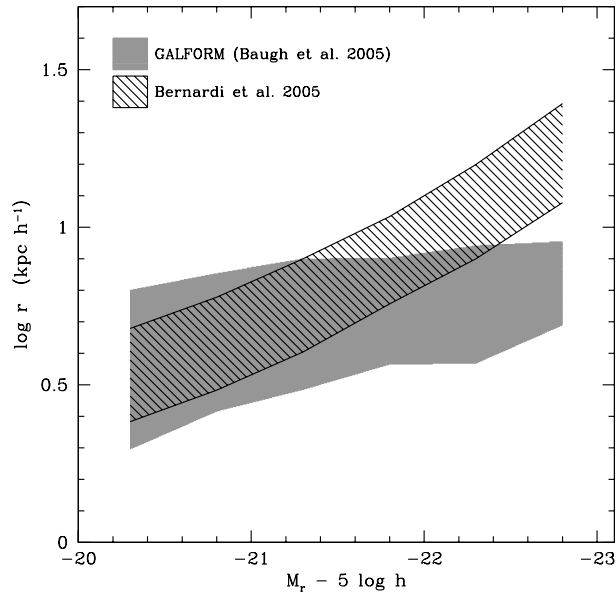}}
\caption{
The relation between half-light radius and r-band magnitude for early-type 
galaxies. Again, the predictions for \g~galaxies are shown in gray and 
the hatched shaded distribution represents the SDSS sample; in both 
cases the shading shows the 10 to 90 percentile range.}
\label{fig:radii}
\end{figure}
\begin{figure}
{\epsfxsize=8.5truecm
\epsfbox[18 144 592 718]{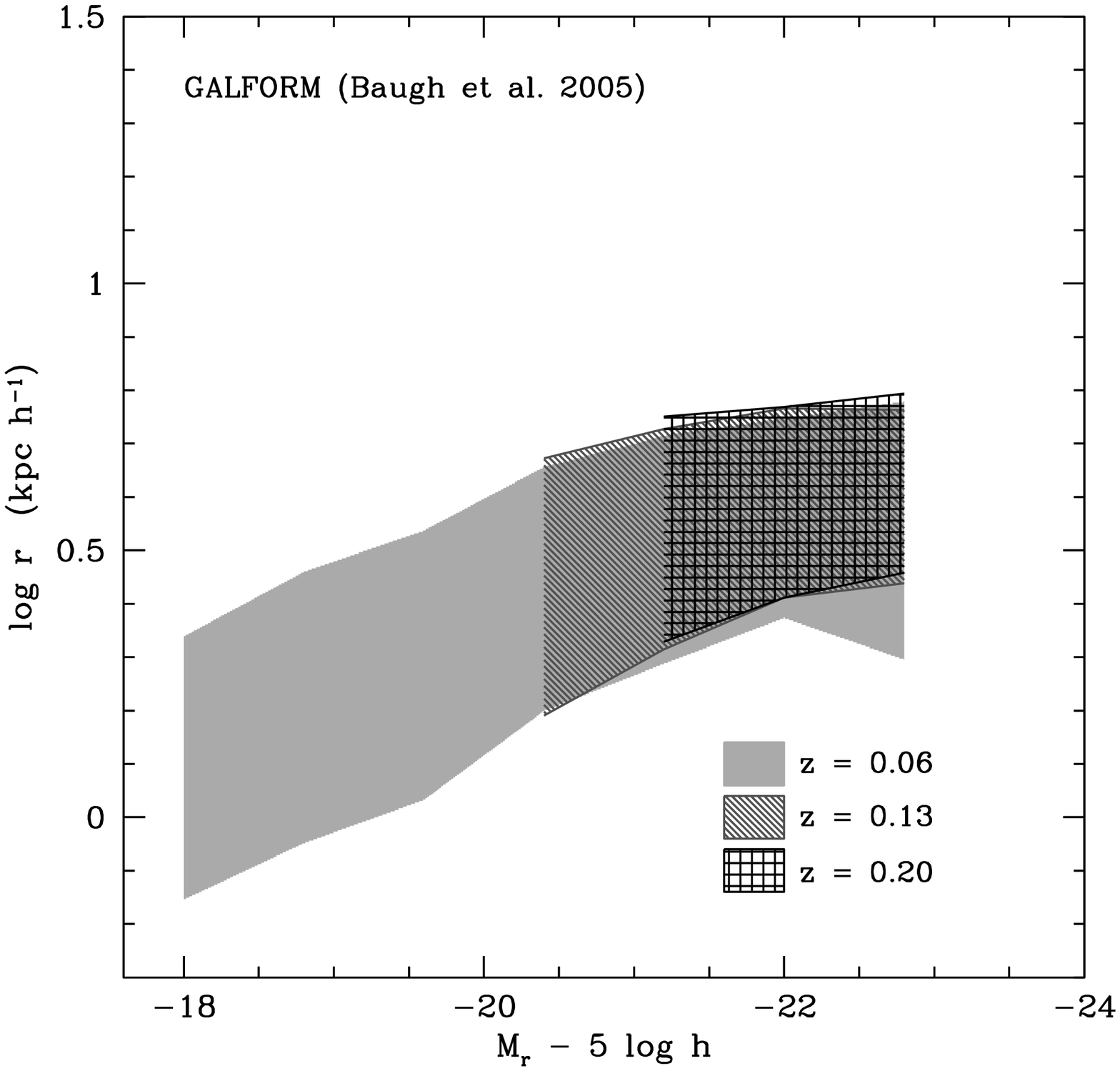}}
{\epsfxsize=8.5truecm
\epsfbox[18 144 592 718]{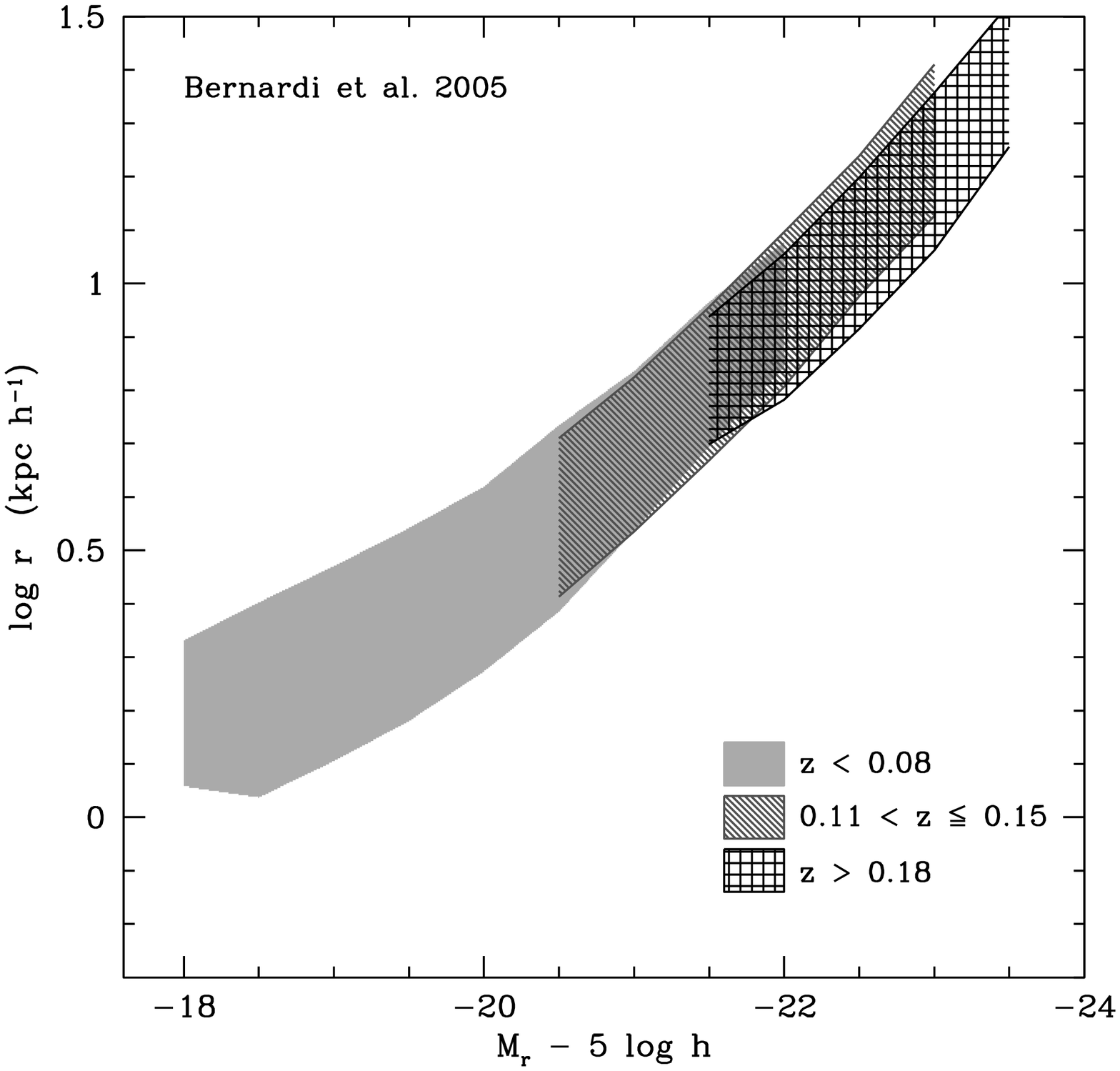}}
\caption{The evolution with redshift of the radius -- r-band luminosity 
relation. The upper panel shows \g~galaxies at the median redshifts of the 
observational subsamples: $z=0.06$, $z=0.13$ and $z=0.20$. The lower panel 
shows the results for the \citet{b05} sample divided into volume-limited bins, 
as indicated by the legend.}
\label{fig:radii.evo}
\end{figure}
We compare the predicted radius-luminosity relation with the Ber05 data 
in Fig.~\ref{fig:radii}. The effective radius plotted here, $r_{\rm e}$, 
is the projected bulge half-light radius of the de Vaucouleurs law, which 
is related to the half-mass radius in 3D, $r_{\rm b}$, 
by:  $r_{\rm e} = r_{\rm b}/1.35$. 
The model predictions do not change if we compute a composite half-mass 
(half-light) radius from the disc and bulge components of the galaxy, as the 
model galaxies we consider are bulge dominated.  
The slope of the predicted radius-luminosity relation is flatter than is observed. 
The agreement between the model predictions and the observations is best 
at fainter magnitudes; the brightest early-type galaxies are predicted to be 
around a factor of three smaller in effective radius than is observed. 
Fig.~\ref{fig:radii} and Fig.~\ref{fig:sigma} suggest that in the model,
the high luminosity early-type galaxies have a pseudo-angular momentum 
which is lower than would be inferred from the data (see \S~\ref{ssec:sizes} for 
the definition of the pseudo-angular momentum of a spheroid).

The evolution of the radius-luminosity relation with redshift is plotted 
in Fig.~\ref{fig:radii.evo}. 
We find no clear change in the slope of the radius-luminosity relation 
over this redshift interval, in agreement with the results of \citet{b03II}. 
We shall return to this point in section~\ref{evolution}.

\subsection{Effective Mass}
We can define an effective dynamical mass, $M_{\rm dyn}$, 
by setting $M_{\rm dyn} \equiv r_e \sigma^2/G$.  
This differs from the true mass, $M$, because the definition of 
$M_{\rm dyn}$ ignores any rotational support and the flattening of galaxies. 
The difference between the two masses can be quantified by a correction 
term, $\xi$: $M = \xi \,M_{\rm dyn}$. For a galaxy with T-type E0, there is no 
flattening or rotational support and so $\xi = 1$. In contrast, for the 
case of an E6 galaxy, the true mass is almost twice as large as the 
dynamical mass, with $\xi \sim 1.9$ \citep[see][for details]{bender}. 

\begin{figure}
{\epsfxsize=8.5truecm
\epsfbox[18 144 592 718]{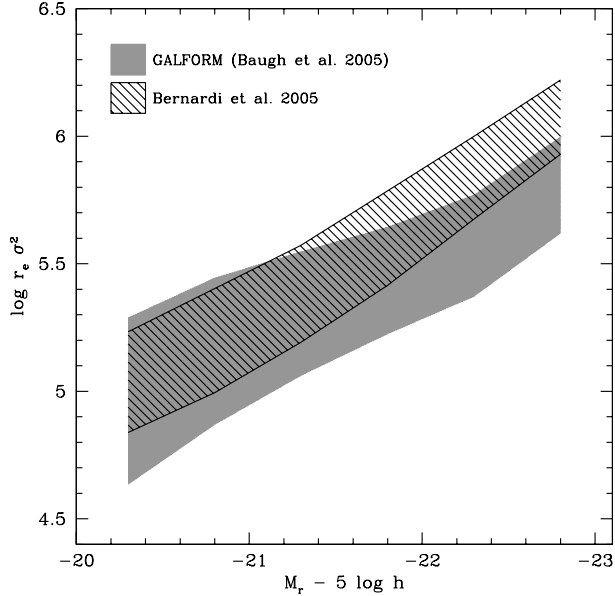}}
\caption{The relation between dynamical mass and luminosity. 
The \g~data is represented in gray and the dark hatched shaded region 
represents the \citet{b05} sample. The shading connects the 10 and 90 
percentile values.}
\label{fig:mass}
\end{figure}

In Fig.~\ref{fig:mass} we compare our prediction for the relation between 
dynamical mass and luminosity with the observed result for the 
\citet{b05} sample. 
The figure reveals reasonable agreement between the model and the 
observations for fainter galaxies. Brighter galaxies, in the model, have a 
somewhat lower dynamical mass than observed. 
As we noted when discussing the radius-luminosity relation, high 
luminosity galaxies in the model display a lower specific pseudo-angular 
momentum than is observed, which translates into a smaller dynamical mass.

\subsection{Fundamental Plane}
\label{sec:fp}
Observational studies indicate that early-type galaxies show 
tight correlations between their kinematic and photometric 
properties \citep{djor,dress}. The remarkably small scatter about 
the so-called fundamental plane connecting the effective radius, 
velocity dispersion and surface brightness of early types encodes 
information about the formation and evolution of these galaxies. 

The existence of a fundamental plane is expected if a stellar system  
obeys the virial theorem, which connects the kinetic and potential 
energies. The assumption of virial equilibrium gives a relation between 
the three-dimensional velocity dispersion, $\sigma_{\rm 3D}$ and 
the ``gravitational'' radius, $r_{\rm g}$, assuming that the system is 
gravitationally bound: 
\begin{eqnarray}
\sigma_{\rm 3D}^2 = \frac{G M}{r_{\rm g}}.
\label{eq:1}
\end{eqnarray}
This equation can be rewritten in terms of the central one dimensional 
velocity dispersion, $\sigma_{\rm 1D}$, and the effective radius of 
the galaxy, $r_{\rm e}$,  
\begin{eqnarray}
\sigma_{\rm 1D}^2 = \frac{G M}{\psi_{\rm r} \psi_{\rm v} r_{\rm e}},
\label{eq:2}
\end{eqnarray} 
where we have defined structural constants such that
\begin{eqnarray*}
\psi_{\rm v} \equiv \frac{\sigma_{\rm 3D}^2}{\sigma_{\rm 1D}^2},\,
\psi_{\rm r} \equiv \frac{r_{\rm g}}{r_{\rm e}},
\end{eqnarray*}
based on the assumption that the population is homologous.

The mean surface brightness within half-mass radius of a galaxy is 
$I_{\rm e} \equiv L/2\pi r_e^2$, where $L$ is the total luminosity of the 
galaxy and the mean surface density is given by 
$\Sigma_{\rm e} \equiv M/2\pi r_{\rm e}^2$. The ratio of the surface brightness 
to the surface density is equal to the mean mass-to-light ratio of the galaxy, within 
$r_{\rm e}$: $\Sigma_{\rm e}/I_{\rm e} = M/L$. Using these 
definitions,  Eq.~\ref{eq:2} can be rearranged to give an expression for 
the fundamental plane, 
\begin{eqnarray}
\nonumber
r_{\rm e} &=& \frac{\psi_{\rm r} \psi_{\rm v}}{2\pi G} \frac{\sigma_{\rm 1D}^2}{I_{\rm e} (M/L)}\mbox{,} \\
\log r_{\rm e} &=& 2 \log \sigma + 0.4 \mu_{\rm e} + \log (\psi_{\rm r} 
\psi_{\rm v}) - \log (M/L) + \gamma, 
\label{eq:3}
\end{eqnarray}
where $\gamma$ is a constant whose value depends upon $G$ and 
the choice of units.

The observed plane is slightly different from the form predicted in 
Eq.~\ref{eq:3}, which follows by applying the virial theorem to a purely 
stellar galaxy without any dark matter and assuming an homologous population. 
For example, \citet{j96} found 
$\log r_{\rm e} = 1.24 \log \sigma + 0.328 \mu_{\rm e} + \gamma'$, 
while \citet{b03III} obtained $\log r_e = (1.49\pm 0.05) 
\log \sigma + (0.30\pm 0.01) \mu_e -(8.78\pm 0.02)$. 

The discrepancy between the theoretical prediction outlined above  
and the observational results is known as the tilt of 
the fundamental plane. 
\cite{trujillo} argued that this tilt is due to a combination of effects: 
structural nonhomology, which means a change in the surface brightness 
profile of the early-types with luminosity, and a variation in the 
mass-to-light ratio of the stellar populations with galaxy luminosity.

The intrinsic thickness or scatter in the fundamental plane poses 
another challenge, and its interpretation is far from clear. 
\citet{forbes98} showed that the scatter was mainly due to the 
age of the stellar population. However, \citet{pahre99} demonstrated 
that the position of the galaxy relative to the FP could not be 
entirely due to age or metallicity effects. 

To determine the location of the fundamental plane, we consider 
an orthogonal fit to the plane given by:
\begin{eqnarray*}
\log  r_{\rm e} = a\,\log\,\sigma + b\,\mu_{\rm e} + c, 
\end{eqnarray*}
and determine the values of the coefficients $a$, $b$ and $c$ 
by minimizing the quantity
\begin{eqnarray*}
\delta = \sum_{i = 1}^{N} \frac{(\log r_{ {\rm e}.i} -  a\,\log\,\sigma_{.i} - b\,\mu_{{\rm e}.i} - c)^2}
{1+a^2+b^2}.
\end{eqnarray*}
Following this procedure, we obtain a fundamental plane for ${\tt GALFORM}$ given by 
$\log  r_{\rm e} = (1.94\pm 0.01) \log \sigma  
+ (0.19\pm 0.01) \mu_{\rm e} - (7.54\pm 0.03)$ in the r-band.

\begin{figure}
{\epsfxsize=8.5truecm
\epsfbox[18 144 592 718]{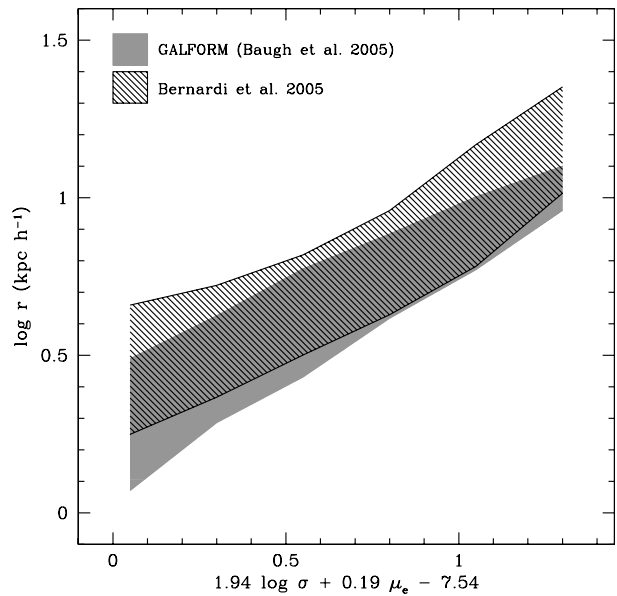}}
\caption{
The fundamental plane for \g~early-type galaxies (gray shading) compared with 
the observational data from Ber05 (hatched shading). The shading denotes 
the 10 to 90 percentile interval. 
}
\label{fig:fp}
\end{figure}

In Fig.~\ref{fig:fp}, we plot the fundamental plane derived from \g~model 
galaxies, along with the data from \citet{b05} in the same projection of the 
plane.
Fig.~\ref{fig:fp} reveals reasonable agreement between the fundamental plane 
predicted by \g~ and the observational data: we can reproduce not 
only the tilt, but also the scatter associated with the plane. In the g-band 
we calculate: $\log  r_{\rm e} = (2.12\pm 0.02) \log \sigma  
+ (0.19\pm 0.01) \mu_{\rm e} - (7.92\pm 0.07)$; and similar results in the 
i and z-bands. This reveals that the slope of the FP is independent of 
wavelength, analogous to the results found by \citet{b03III}.

\begin{figure}
{\epsfxsize=8.5truecm
\epsfbox[18 144 592 718]{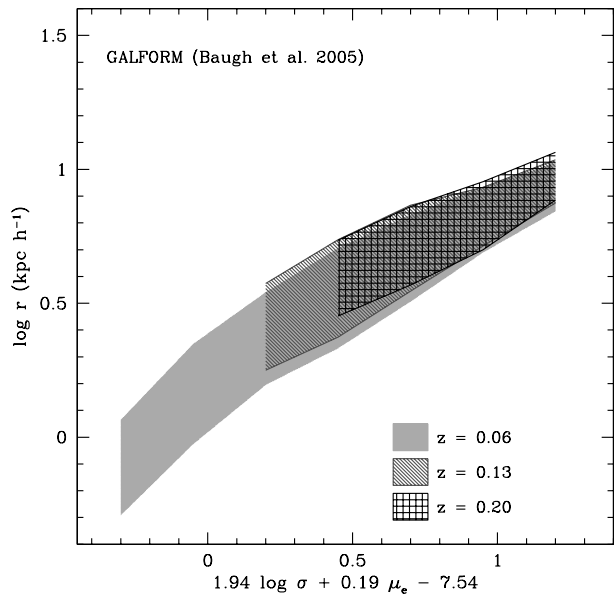}}
{\epsfxsize=8.5truecm
\epsfbox[18 144 592 718]{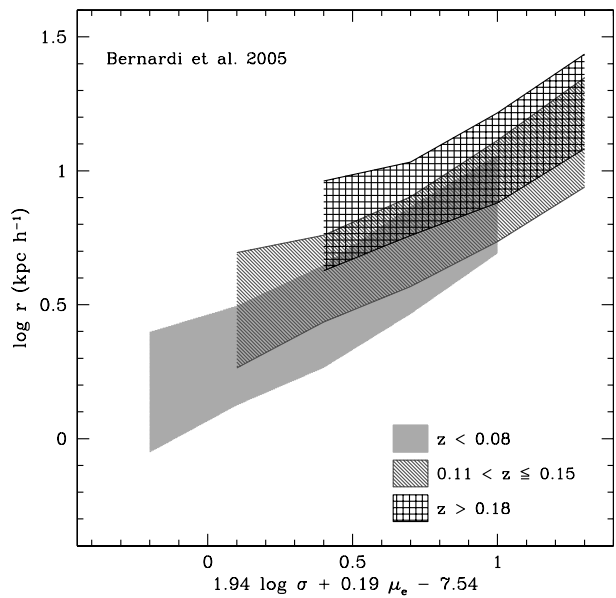}}
\caption{
The evolution of the fundamental plane with redshift. The upper panel shows 
model predictions for redshifts $z=0.06$, $z=0.13$ and $z=0.20$; these 
are the median redshifts of the observational samples plotted in the 
lower panel. Again, the shaded regions show 
the 10 to 90 percentile ranges of the distributions.
}
\label{fig:fp.evo}
\end{figure}

We plot the fundamental plane at different redshifts in 
Fig.~\ref{fig:fp.evo}. The radius, velocity dispersion, surface 
brightness and mass-to-light ratios of the model galaxies all evolve 
with time (see \S 5), so one might expect to see some evolution in 
the fundamental plane itself, unless the changes in these quantities 
occur in such a way as to cancel out any evolution in the locus of 
the plane. 
Ber05 report evolution in the position of the fundamental plane 
which corresponds to a change in the mean galaxy surface brightness 
of $\Delta \mu_{\rm e} \approx -2 z$. We find no 
clear evidence for evolution in the model predictions over the 
same redshift interval. In section~\ref{evolution} we show the 
predictions for the scaling relations over a wider baseline in 
redshift.

\null

\section{The dependence of the structural properties of ellipticals 
on the physical ingredients of the model}
\label{section:changes}

Our calculation of the sizes of galactic spheroids contains several 
steps and is sensitive to some of the physical ingredients of the 
galaxy formation model more than others. The beauty of 
semi-analytical modelling is that we can switch off or vary 
particular assumptions or processes to isolate their 
impact on the model predictions. Such a study is only possible to a very 
limited extent in fully numerical simulations of galaxy formation. Moreover, 
the high speed of the semi-analytical calculations compared with a numerical 
simulation allows us to examine many different variants in a short time. 
In this section, we seek to establish the sensitivity of our model predictions 
for the structural and photometric properties of spheroids to  
the composition of the model. For this purpose, we study the model 
predictions at $z=0$ and consider bulge dominated galaxies, i.e. those 
with a bulge-to-total luminosity ratio in the $r$-band of $B/T > 0.8$. 
The results of this section are presented in 
Figs.~\ref{fig:delta.fp},~\ref{fig:sigma.changes},~\ref{fig:radii.changes} 
and~\ref{fig:fp.changes}, which look, respectively, at how deviations 
from the fundamental plane correlate with various galaxy properties, 
the Faber-Jackson relation between velocity dispersion and luminosity, 
the radius-luminosity relation and the fundamental plane. 

\subsection{The deviation from the fundamental plane}

As we noted in the previous section, there is some controversy in 
the literature regarding the source of the dispersion around 
the fundamental plane, with some authors arguing that the 
scatter could be due to a number of causes, such as variations in 
the formation times of galaxies, metallicity trends in stellar 
populations, or differences in the dark matter content of galaxies.

\begin{figure*}
{\epsfxsize=14.5truecm
\epsfbox[18 144 592 718]{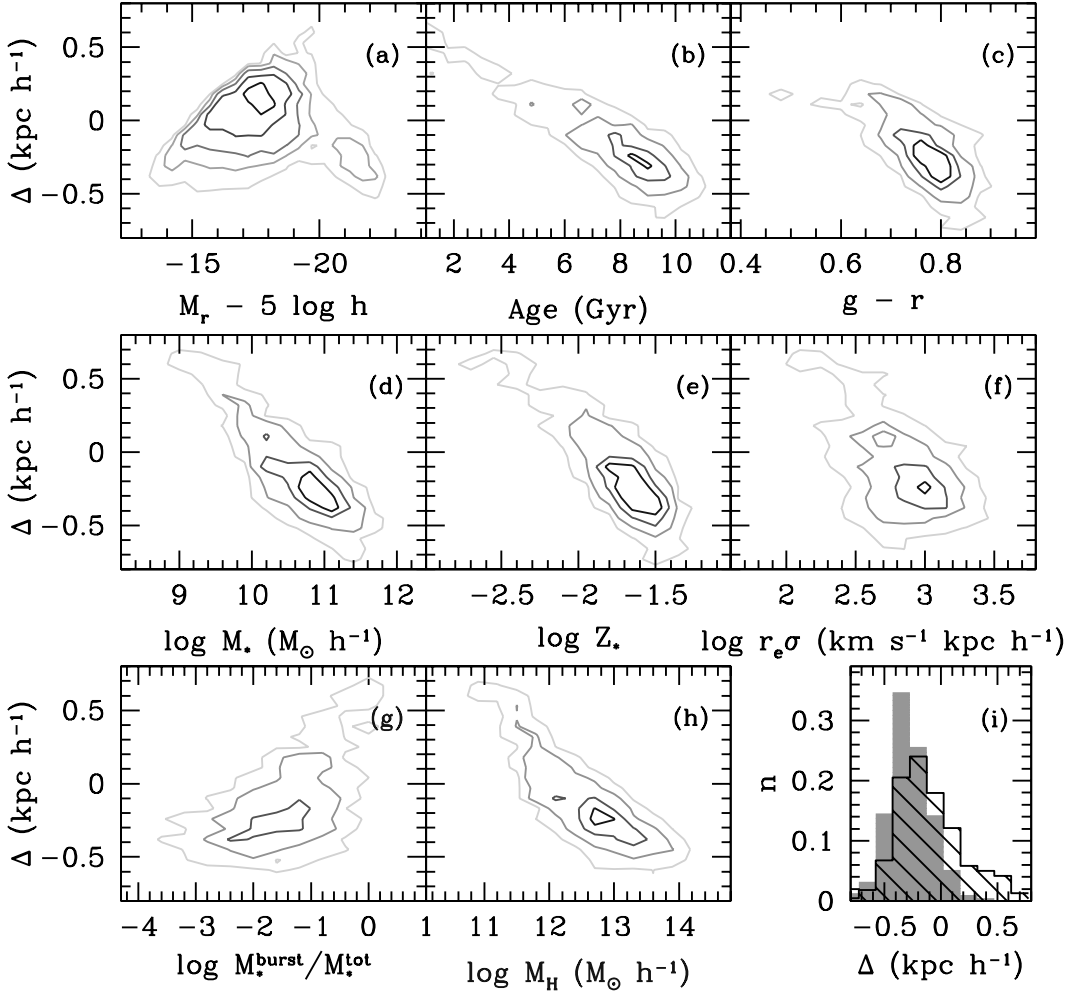}}
\caption{
The dependence of the deviation (defined as 
$\Delta = \log r_e - (1.94 \log \sigma + 0.19 \mu_e - 7.54$)) 
from the fundamental plane in the model on various galaxy properties: 
(a) the $r$-band absolute magnitude; then, for $M_r-5\log h<-19.5$ galaxies 
(b) the $r$-band luminosity-weighted age, 
(c) the $g - r$ colour, 
(d) the total stellar mass, 
(e) the stellar metallicity,
(f) the pseudo-specific bulge angular momentum,  
(g) the ratio between the mass of the stars formed in the last burst, $M_*^{\rm burst}$, 
and the total stellar mass, $M_*^{\rm tot}$, at the present day, 
(h) the halo mass; and 
(i) whether or not the galaxy is a central galaxy or a satellite.
The contours are indicative of the density of model galaxies. 
In the panel (i), the hatched histogram represents the deviation from 
the fundamental plane for the central galaxies, and the grey histogram 
shows the distribution for the satellite galaxies.
}
\label{fig:delta.fp}
\end{figure*}

Fig.~\ref{fig:delta.fp} shows how the deviation from the fundamental 
plane correlates with various galaxy properties.
Here, the quantity $\Delta$ represents the offset from the $z=0$ 
fundamental plane predicted in the $r$-band, after applying 
the selection criteria to match the Ber05 SDSS sample: 
$\Delta = \log r_b - (1.94 \log \sigma + 0.19 \mu_e - 7.54)$. 
Fig.~\ref{fig:delta.fp}(a) shows that the 
deviation is correlated with r-band absolute magnitude for 
galaxies fainter than $L_*$. This result shows that magnitude-limiting 
a sample might bias the determination of the fundamental plane. 
Surprisingly, luminous early-type galaxies ($M_r-5 \log h<-20$) 
exhibit no correlation with deviation from the fundamental 
plane, which is in agreement with the results of \citet{b03II}. 
For the remaining panels in Fig.~\ref{fig:delta.fp}, we 
only select galaxies with $M_r -5 \log h<-19.5$ (i.e. brighter than 
one magnitude faintwards of $M_*$), 
in order to make our results comparable to observations.
Fig.~\ref{fig:delta.fp}(b) reveals a strong anticorrelation 
between the deviation and the r-band luminosity-weighted age 
of the galaxy, in the sense that galaxies which lie above 
the fundamental plane are younger. A linear fit to the 
distribution reveals $\Delta = -(0.11\pm 0.03)$ Age 
$+\,(0.64\pm 0.09)$. This strong correlation 
indicates that the age of the stellar population 
plays an important role in determining the position of the 
galaxy in the fundamental plane space 
\citep[see][]{forbes98, pahre99}. 
We find that the fundamental plane offset is also anticorrelated 
with g-r colour (Fig.~\ref{fig:delta.fp}(c)). As noted by 
\citet{b03III}, this is due to the correlation 
between colour and velocity dispersion 
(see also Bernardi et~al. 2005). 
Interestingly, we see in Fig.~\ref{fig:delta.fp}(d) 
that the total stellar mass is anticorrelated with the 
deviation from the fundamental plane, for galaxies 
brighter that $M_r-5\log h<-19.5$. When all the early-type 
galaxies are included, then the distribution reveals a different 
picture: similar to the trend seen in panel (a), we find that 
faint galaxies show a stronger deviation from the fundamental plane. 
Fig.~\ref{fig:delta.fp}(e) shows that the absolute 
metallicity of the stellar population is anticorrelated 
with a deviation from the fundamental plane: metal-rich 
galaxies are to be found predominately below the mean 
fundamental plane relation. The relation between 
the pseudo-specific angular momentum, $r_{\rm e}\sigma$, 
of the bulge and the FP offset is plotted in Fig.~\ref{fig:delta.fp}(f), 
revealing a weak anticorrelation between these quantities. 
There is little correlation between the deviation from 
the fundamental plane and the fraction of the total stellar 
mass formed in the last burst of star formation triggered 
by a galaxy merger (Fig.~\ref{fig:delta.fp}(g)), which 
shows that the presence of gas in galaxy mergers does not 
change significantly the fundamental plane relation. 
This seems to contradict the recent results of \citet{rob}. 
However, if we also consider faint galaxies, which 
tend to have mergers containing a larger fraction of gas, then 
we find an anticorrelation between $M_*^{\rm burst}/M_*^{\rm tot}$ 
and the deviation from the FP, along the lines of that seen by 
Robertson et~al. 
The relation between the deviation from the FP and halo mass is 
shown in Fig.~\ref{fig:delta.fp}(h). We find an anticorrelation 
between these two quantities, such that galaxies which lie in 
more massive haloes are found below the main FP relation, 
i.e. cluster galaxies should lie below the mean fundamental 
plane. In Fig.~\ref{fig:delta.fp}(i), we show that the 
distribution of the FP offset for central galaxies 
resembles that predicted for satellite early-types.

\subsection{The physics of the model and the scaling relations}
\begin{figure*}
{\epsfxsize=14.5truecm
\epsfbox[18 144 592 718]{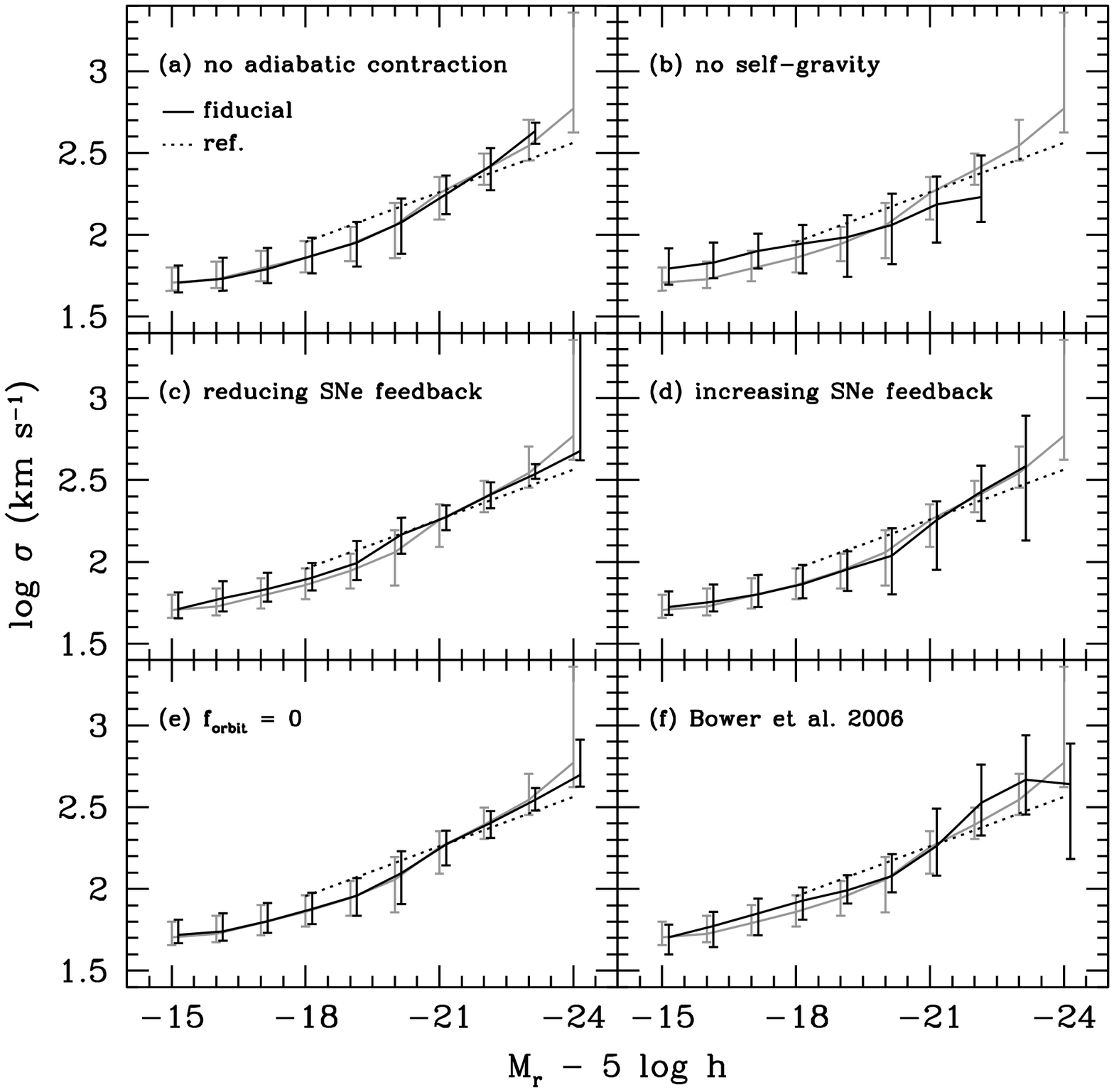}}
\caption{
The sensitivity of the model predictions for the 
Faber-Jackson relation to: 
(a) switching off adiabatic contraction, 
(b) omitting the self-gravity of the baryons, 
(c) reducing the strength of supernova feedback,  
(d) increasing the strength of SNe feedback, 
(e) the omission of the orbital energy from the calculation of 
the size of the merger remnant 
and 
(f) using the \citet{bower} model with AGN feedback. 
In each panel, the grey line shows the median prediction from the 
reference model \citep{baugh05}, at $z=0$. The black solid line shows 
the median for the variant model. 
The errorbars indicate the 10 to 90 percentile of the predictions.
The dotted line in each panel shows the observed 
relation for SDSS early-type galaxies for reference \citep{b03II}. 
}
\label{fig:sigma.changes}
\end{figure*}

In this section, we examine how the predictions of the Baugh et~al. 
model change if one ingredient at a time is varied. These variant 
models are not necessarily acceptable galaxy formation models, because 
they may not give as good a match to the local data used to calibrate 
the model parameters as was the case for the fiducial Baugh et~al. model. 
We also show the predictions of the Bower et~al. model, as a further 
example of a variant model. In this case, many ingredients have been 
changed from the ones used in the Baugh et~al. model, as explained at 
length in Section 2. 

\begin{figure*}
{\epsfxsize=14.5truecm
\epsfbox[18 144 592 718]{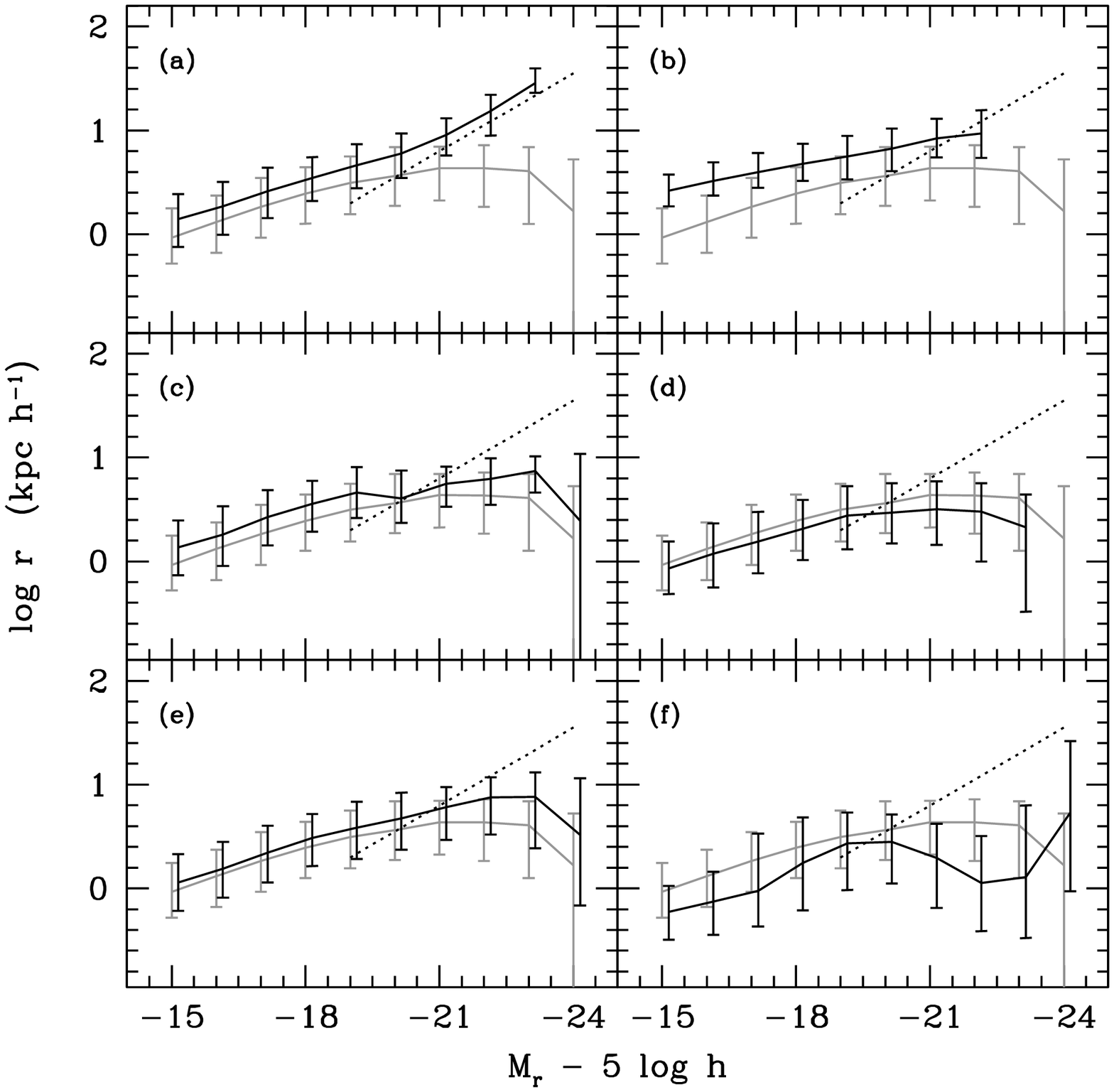}}
\caption{
The sensitivity of the relation between radius and luminosity to: 
(a) switching off adiabatic contraction, 
(b) omitting the self-gravity of the baryons, 
(c) reducing the strength of supernova feedback,  
(d) increasing the strength of SNe feedback, 
(e) the omission of the orbital energy from the calculation of 
the size of the merger remnant 
and 
(f) using the \citet{bower} model with AGN feedback. 
In each panel, the grey line shows the median prediction from the 
reference model \citep{baugh05}, at $z=0$. The black solid line shows 
the median for the variant model. 
The errorbars indicate the 10 to 90 percentile of the predictions.
The dotted line in each panel shows the observed 
relation for SDSS early-type galaxies for reference \citep{b03II}. 
}
\label{fig:radii.changes}
\end{figure*}

The first ingredient we test is the adiabatic contraction model 
used to take into account the gravitational pull of the baryons on 
the dark matter. 
To recap, the condensation of baryons at the centre of the dark matter 
halo provides an additional gravitational force on the dark matter which 
causes it to move inwards, thereby increasing the density of dark matter 
in the central part of the halo. This in turn alters the gravitational 
force on the baryons due to the dark matter. The degree of contraction is 
computed by exploiting the fact that, in a slowly varying potential, 
the action integral, $\oint p_i dq_i$, is an adiabatic invariant 
for each particle of mass $i$, where $p_i$ is the conjugate momentum of 
the coordinate $q_i$ \citep{barnes,blumenthal,jesseit}. If we assume spherical 
symmetry and circular orbits, the action integral simplifies to the 
conservation of angular momentum in spherical shells, $r M(r)$. The adiabatic 
contraction of the dark matter leads to a more centrally peaked halo 
density.
The main consequence of switching off the adiabatic contraction of the 
dark matter halo is that the half-mass radius of the spheroid increases 
(Fig.~\ref{fig:radii.changes}(a)). The radii of bright galaxies increase 
by a larger factor than those of faint galaxies, leading to a steepening 
of the radius-luminosity relation. The slope of the radius-luminosity is 
in much better agreement with the observed slope on omitting adiabatic 
contraction, although the model galaxies are too large overall (both 
spheroids and discs). 
\begin{figure*}
{\epsfxsize=14.5truecm
\epsfbox[18 144 592 718]{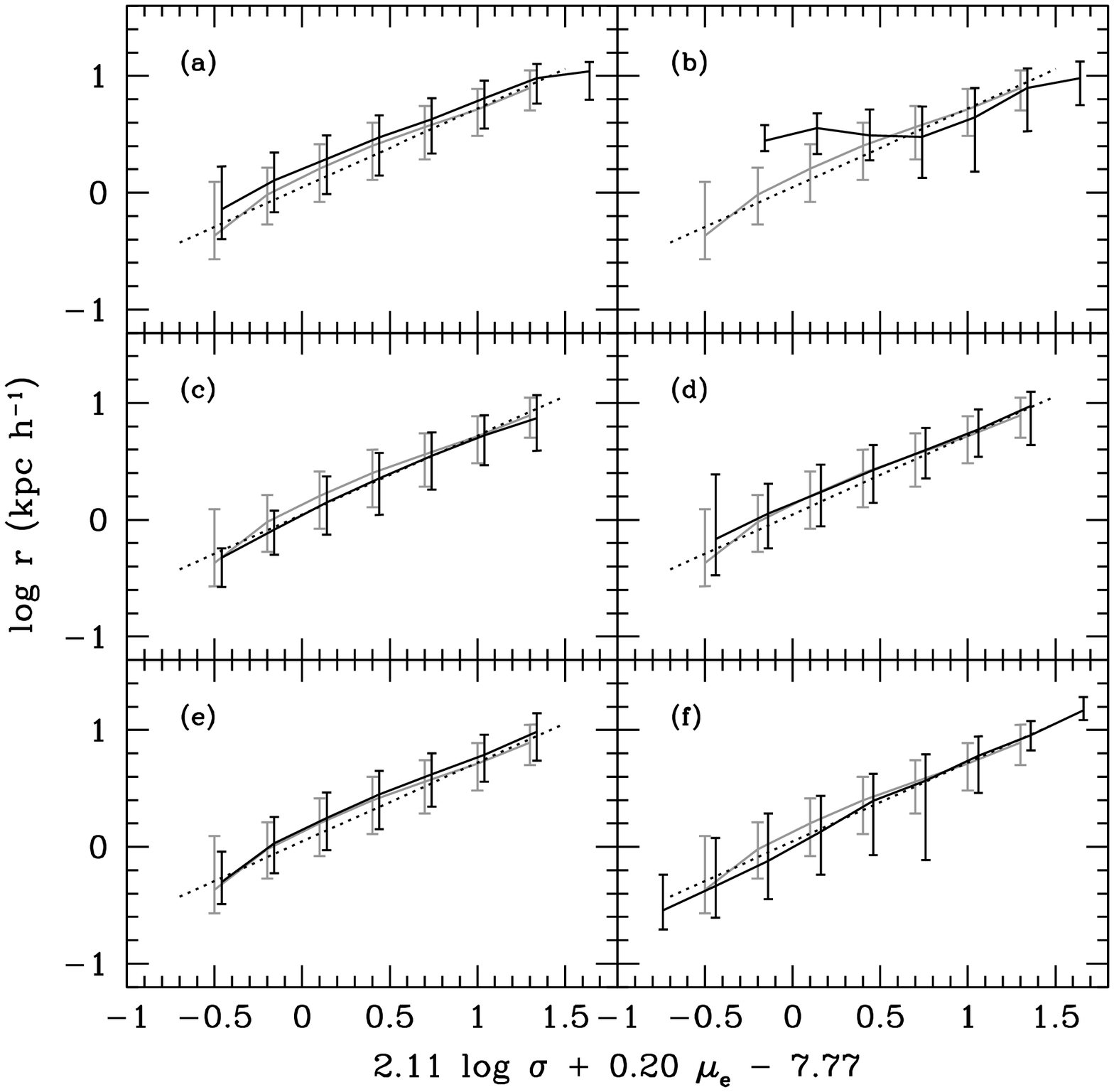}}
\caption{
The sensitivity of the fundamental plane to: 
(a) switching off adiabatic contraction, 
(b) omitting the self-gravity of the baryons, 
(c) reducing the strength of supernova feedback,  
(d) increasing the strength of SNe feedback, 
(e) the omission of the orbital energy from the calculation of 
the size of the merger remnant 
and 
(f) using the \citet{bower} model with AGN feedback. 
In each panel, the grey line shows the median prediction from the 
reference model \citep{baugh05}, at $z=0$. The black solid line shows 
the median for the variant model. 
The errorbars indicate the 10 to 90 percentile of the predictions.
The dotted line in each panel shows the observed 
relation for SDSS early-type galaxies for reference \citep{b05}. 
}
\label{fig:fp.changes}
\end{figure*}

Next we ignore the self-gravity of the baryons when computing the size and 
effective rotation speed of the disc and bulge. This also means that there 
is no adiabatic contraction. The rotation curve of the galaxy in this 
case is set purely by the dark matter, which is assumed to have an NFW 
density profile \citep{nfw}. 
The consequences of this change are a flattening in the velocity 
dispersion-luminosity relation (Fig.~\ref{fig:sigma.changes}(b)), with brighter 
galaxies displaying a lower velocity dispersion, and a uniform 
increase in the radius of the spheroid (Fig.~\ref{fig:radii.changes}(b)). 
In combination, these changes result in a different projection of the 
fundamental plane which looks flatter in the projection which best fits 
the predictions of the fiducial model.

Feedback, the regulation of the star formation rate due to the reheating and 
ejection of cooled gas following the injection of energy into the interstellar 
medium by supernova explosions, plays an important role in setting the 
sizes of disc galaxies \citep{cole00}. 
The strength of SNe feedback is quantified in the model by the 
parameter $\beta$ \citep[for details see][]{cole00}: 
$\beta = (V_{\rm hot}/V_{\rm disc})^{\alpha_{\rm hot}}$, where 
$V_{\rm hot}$ and $ \alpha_{\rm hot}$ are parameters and $V_{\rm disc}$ is 
the rotation speed of the disc at the half mass radius. 
The mass of cold gas which is reheated is given by  
$\dot{M}_{\rm reheat} = \beta \psi$, where $\psi$ is the star formation rate. 
In the Baugh et~al. model, the values adopted for these parameters are: 
$\alpha_{\rm hot} = 2$ and $V_{\rm hot} = 300\,{\rm km\,s}^{-1}$. 
We show the impact of reducing (by setting $v_{\rm hot}=100\,{\rm km \,s}^{-1}$) 
and increasing (by setting $v_{\rm hot}=600\,{\rm km\, s}^{-1}$) 
the strength of supernova feedback in Figs.~\ref{fig:sigma.changes},
~\ref{fig:radii.changes} and~\ref{fig:fp.changes}-(c)-(d).
Cole et~al. demonstrated that increasing the strength of supernova feedback 
results in gas cooling to form stars in larger haloes, which leads to larger 
discs. Conversely, reducing the feedback allows gas to cool and form stars 
in smaller haloes resulting in smaller discs. These trends are reproduced in 
Figs.~\ref{fig:radii.changes}(c) and (d). 
There is little change in velocity dispersion on changing the strength of
the supernova feedback. The shift in the zero-point of the radius-luminosity 
relation produces a change in the location of the fundamental plane (see 
Fig.~\ref{fig:fp.changes}-(c)-(d)).

In the fiducial \g~model, spheroids are the end products of galaxy mergers. 
As we explained in \S~\ref{section:model}, the radius of the merger remnant 
is determined by conserving the binding energy of the individual galaxies 
involved in the merger and their relative orbital energy. The contribution 
of the orbital energy to the energy budget is parameterized by 
$f_{\rm orbit}$: the standard choice is to set $f_{\rm orbit}=1$ and to 
include the full orbital energy in the calculation of the remnant size. 
In Figs.~\ref{fig:sigma.changes}, ~\ref{fig:radii.changes} 
and~\ref{fig:fp.changes}-(e), we show the effect of removing the 
contribution of the orbital energy from the calculation of the radius of 
the spheroid produced by mergers, i.e. we set $f_{\rm orbit} = 0$. 
Perhaps surprisingly, this change results in an imperceptibly small 
change in the radius of the spheroid, except in the case of the brightest 
galaxies.

Finally, we consider the model of \citet{bower}, who implemented an AGN 
feedback scheme into {\tt GALFORM}, in which cooling flows are quenched 
in massive haloes at low redshift. As a result of this change to the 
cooling model in \g, \citet{bower} were able to produce improved 
matches to the  
local B and K-band luminosity functions, the observed bimodality of colour 
distribution and the inferred evolution of the stellar mass function. 
In Figs.~\ref{fig:sigma.changes}, ~\ref{fig:radii.changes} 
and~\ref{fig:fp.changes}-(f) we plot the scaling relations for 
\citet{bower} model. Though the model performs quite well 
in reproducing the local fundamental plane of early-type galaxies 
and the Faber-Jackson relation, the radius-luminosity relation 
for bright galaxies is substantially different from both the observations 
and from the predictions of the \citet{baugh05} model: luminous 
galaxies in the Bower et~al. model are up a factor of three smaller 
in radius than in the Baugh et~al. model. 

We also considered a variant of the Baugh et~al. model in which 
the Kennicutt IMF was used in starbursts, in place of the top-heavy 
IMF. This produces scaling relations for early type galaxies which 
look very similar to those presented for the Bower et~al. model 
in Figs~\ref{fig:sigma.changes}~\ref{fig:radii.changes}, 
and ~\ref{fig:fp.changes}, with the main change being a shift in 
the predicted radius-luminosity relation. When the IMF is changed, 
the yield and recycled gas fraction are also changed accordingly, 
which affect the rate at which gas cools and alter the star formation 
timescale. This suggests that the primary difference in the predictions 
of the Baugh et~al. and Bower et~al. models is due to the choice of 
the IMF used in starbursts, in spite of the other differences between 
the models outlined in Section 2. 

The results in this section suggest that the scaling relations 
of early-type galaxies are essentially insensitive to variations 
in some of the model parameters. However, this should not be 
interpreted as implying that these observations are of limited 
value in constraining the models. It should be remembered that 
our starting point is a model of galaxy formation which has 
already successfully passed a range of comparisons with observed 
galaxy properties.

\section{The evolution of scaling relations}
\label{evolution}
We now present the \g~predictions for the evolution of the structural 
and photometric properties of early-type galaxies with redshift. 
In this section, we consider the evolution over a much wider baseline 
in redshift than we addressed in the previous section. 
Furthermore, in order to get a clear picture of the nature of the 
evolution, we relax some of the selection criteria which we applied 
to the model output in previous sections, where the goal was 
to mimic the Ber05 sample selection as closely as possible. 
The only selection we apply in this section is that the bulge 
must account for at least 80\% of the total luminosity in the 
rest-frame $B$-band. 

\begin{figure}
{\epsfxsize=8.5truecm
\epsfbox[18 144 592 718]{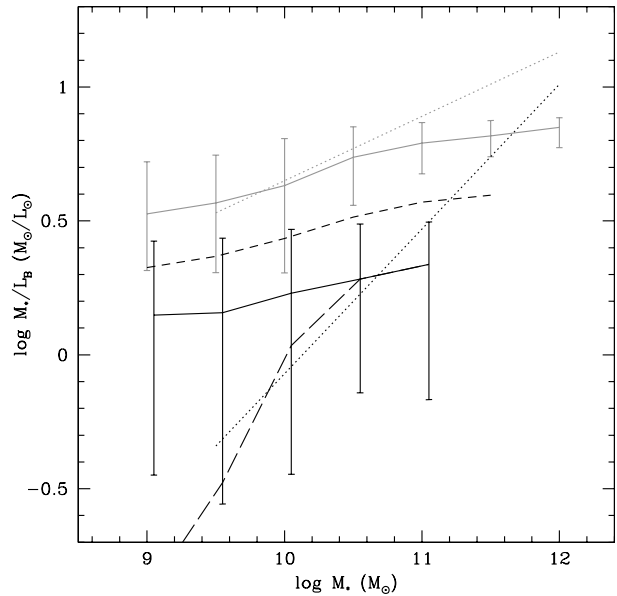}}
\caption{
The predicted evolution with redshift of the mass-to-light ratio 
in the rest-frame $B$-band, plotted against stellar mass. The 
grey line shows the prediction for $z=0$, the short 
dashed line for $z=0.5$, the solid black line for $z=1$. 
The long dashed line shows the prediction for the median 
mass-to-light ratio at $z=1$, when only considering galaxies 
with $M_{\rm B}-5 \log h < -19.5$.
The dotted lines show the relations found by \citet{j06} 
for the Coma cluster (grey), $z\approx 0$, and a high redshift sample 
(black), $z\approx 1$.
}
\label{fig:ml.z}
\end{figure}
It is important to be able to disentangle changes in the typical 
stellar populations of early-type galaxies with redshift from 
evolution in their structural properties. Hence, we first examine 
the predicted evolution in the mass-to-light ratio of early-type 
galaxies in Fig.~\ref{fig:ml.z}. The stellar populations of 
early-type galaxies at $z=1$ in \g~have lower mass-to-light 
ratios by a factor of $\approx 3$ compared with the early-types 
at $z=0$. This result is in agreement with the change in 
mass-to-light ratio inferred from observations by 
\citet{ven}, $d(\log M/L)/dz = -0.47\pm 0.11$ 
\citep[see also][]{dokkum06, wel, treu}. We find a weak 
dependence of mass-to-light ratio on total stellar mass. 
However this is somewhat lower than what is seen 
observationally \citep[see][]{wuyts, wel}. Note that when 
we restrict our attention to bright galaxies (i.e. those
with $M_{\rm B}-5 \log h <-19.5$), the prediction for the 
median mass-to-light ratio steepens considerably, bringing 
the model predictions into much better
agreement with the observational estimates.

\begin{figure}
{\epsfxsize=8.5truecm
\epsfbox[18 144 592 718]{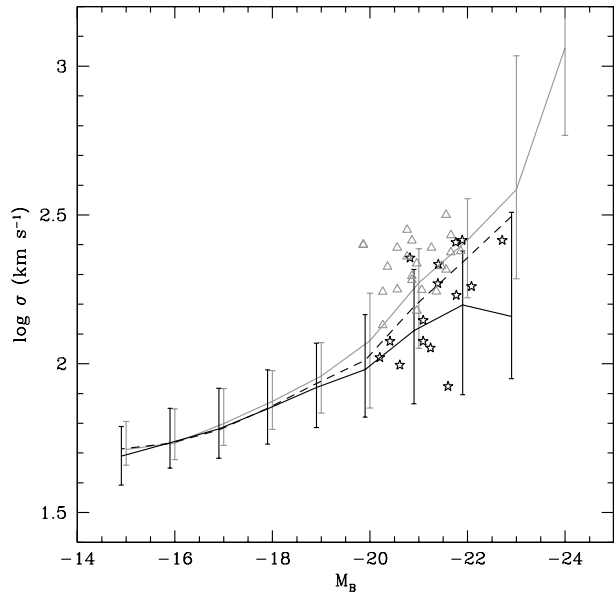}}
\caption{
The predicted evolution with redshift of the relation between 
velocity dispersion and rest-frame $B$-band luminosity. 
The grey solid line shows the Faber-Jackson relation in the local 
universe ($z=0$), while the short-dashed line shows it at $z=0.5$ and 
the black solid line at $z=1$. The errorbars show the 10 to 90 
percentile range of the distribution. The gray triangles show 
local data from \citet{reda} and the stars represent the 
sample of early-type galaxies in the K20 survey by 
\citet{serego}, with $z\approx 1$.
}
\label{fig:sigma.z}
\end{figure}

The evolution with redshift of the Faber-Jackson relation is shown 
in Fig.~\ref{fig:sigma.z}. We show the correlation between 
velocity dispersion and the $B$-band magnitude in the rest-frame 
for the local universe ($z=0$), $z=0.5$ and $z=1$. 
Fig.~\ref{fig:sigma.z} shows that the model predicts differential 
evolution in velocity dispersion with rest-frame luminosity; at 
brighter luminosities, the velocity dispersion drops by up to a factor 
of $\approx 3$ between $z=0$ and $z=1$, whereas for fainter luminosities, 
the change in velocity dispersion is much more modest. These results 
are similar to the ones found in observational studies 
\citep[cf.][]{serego, reda}. We note that the scatter around 
the FJ relation seems to increase slightly with redshift. 
Furthermore, it is clear that the slope of 
the relation for faint early-type galaxies is shallower 
than that of the bright-end, which resembles the results for faint 
galaxies found by some authors \citep[e.g.][]{mac, davies}.

\begin{figure}
{\epsfxsize=8.5truecm
\epsfbox[18 144 592 718]{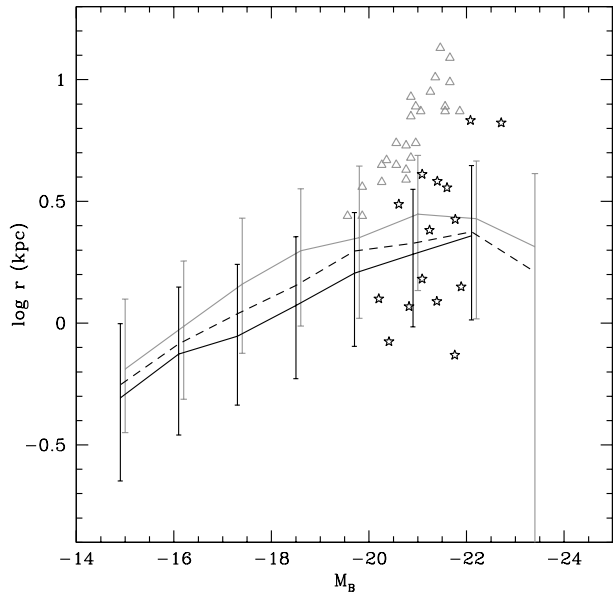}}
\caption{
The predicted evolution with redshift of the relation between 
radius and luminosity. The grey solid line shows the relation 
in the local universe, the black dashed  line at $z=0.5$ and 
the black solid line at $z=1$. The errorbars show the 10 to 90 
percentile range of the model predictions. The triangles show 
local data from \citet{reda} and the stars are galaxies in 
the K20 survey by \citet{serego}, with $z\approx 1$.}
\label{fig:radii.z}
\end{figure}

Fig.~\ref{fig:radii.z} shows how the relation between radius 
and luminosity varies with redshift.  As we saw in Section 4, 
the model predicts that the brightest early-types are too 
small. Nevertheless, it is interesting to look at the 
predictions at different redshifts. 
The primary agent behind the shift in the predictions is the passive 
evolution of the stellar populations in the elliptical galaxies. 
Also, at a fixed stellar mass, the galaxy radii decrease with 
redshift (Coenda et~al 2007; in preparation).
As redshift increases, we see the bulk of the stars in ellipticals 
when they were younger and hence brighter. There is no significant 
trend in the size of the scatter in this relation with redshift. 

\citet{trujillo_b} estimated the evolution of the radius-luminosity 
relation of bright galaxies up to $z\sim 3$ \citep[see also][]{bouw}. 
These authors found that early-type galaxies, as defined by a 
high value of S\'ersic index, show a size evolution proportional to 
$(1+z)^{-1.01\pm 0.08}$; this is comparable to the amount of 
evolution we predict in Fig.~\ref{fig:radii.z} for all ellipticals.

The evolution of the fundamental plane has long been used to 
study changes in the stellar populations of galaxies 
\citep[e.g.][]{dokkum96, dokkum01, gebhardt, wel06}.
As previously noted, \citet{b03III} found evolution in 
the fundamental plane which is consistent with the passive 
aging of the stellar population, $\Delta \mu_e \approx -2 z$, 
but without any noticeable difference in the slope. 
There is a general consensus in the literature regarding 
the nature of the evolution of the fundamental plane 
\citep{gebhardt, ziegler, j06}. However, as we have shown in 
the previous section, deviations from the fundamental 
plane relation are linked to several galaxy properties.

\begin{figure}
{\epsfxsize=8.5truecm
\epsfbox[18 144 592 718]{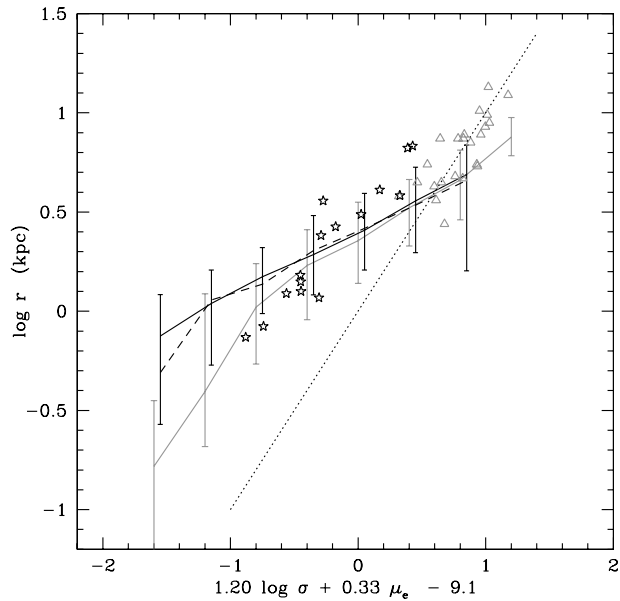}}
\caption{
The predicted evolution of the fundamental plane with redshift 
for galaxies brighter than $M_{\rm B}-5 \log h < -19.5$. 
The grey solid line shows the relation in the local universe, 
the dashed black line at $z=0.5$ and the black solid line at $z=1$. 
The errorbars show the 10 to 90 percentile range of the model 
predictions. The dotted line shows the relation found by 
\citet{j96}. The gray triangles represent $z=0$ data 
from \citet{reda} and the stars show data from \citet{serego}, 
at $z\approx 1$.
}
\label{fig:fp.z}
\end{figure}
In Fig.~\ref{fig:fp.z}, we plot the model predictions in the \citet{j06} 
projection of the fundamental plane, 
$\log  r_e = 1.2 \log \sigma + 0.33 \mu_e - 9.1$, 
for galaxies with $M_{\rm B}-5 \log h < -19.5$. 
In this projection, we find no evolution in the slope or offset 
of the fundamental plane up to $z=1$, for galaxies with 
$\log r_{\rm e}\ga 0.3$, which is contrary to the claims made from 
observations. As we noted in Fig.~\ref{fig:delta.fp}(a), the evolution 
found in observational studies might be partly due to the 
correlation between the magnitude and the deviation 
from fundamental plane, i.e. magnitude-limiting samples 
might induce the zero-point of the FP to shift. On 
the other hand, the effective radius of bright early-type 
galaxies in \g~is smaller than observed (see Fig.~\ref{fig:radii}), 
and evolves with redshift, which will complicate any inferences drawn 
from the evolution of the FP. 
Interestingly, in the small radius regime, the evolution of 
our predicted fundamental plane shows an offset similar 
to that observed.

\section{Discussion and conclusions} 

We presented tests of the model proposed by \citet{cole00} to 
calculate the scale sizes of the disc and bulge components of 
galaxies. This is currently the most sophisticated model in use in 
semi-analytical codes to compute the radii of galaxies. In brief, 
the model assumes that galactic discs have an exponential profile and 
that spheroids follow an $r^{1/4}$ law in projection. The hot gas 
atmosphere in dark matter haloes is assumed to have the same specific 
angular momentum as the dark matter. Gas is assumed to retain its 
angular momentum as it cools to form a galactic disc. The size of a 
merger remnant is computed by conserving the sum of the binding and 
orbital energies of the merging galaxies and applying the virial 
theorem. The self-gravity of the baryons and their impact on the 
distribution of dark matter in the central parts of the halo are taken 
into account. Cole et~al. demonstrated that this 
model predicts  scale length distributions for galactic discs which are 
in excellent agreement with observations. 

In this paper, we have carried out the first tests of the model predictions 
for the structural properties of early-type galaxies and the evolution 
of these relations with redshift, using the published models of 
\citet{baugh05} and \citet{bower}. 
The Baugh et~al. and Bower et~al. models differ in a number of ways, as 
set out in Section 2. Two of the main differences are the manner in which 
the models prevent the overproduction of bright galaxies and in the IMF 
assumed in starbursts. Bower et~al. use AGN heating to switch off the 
cooling-flow in haloes with a quasistatic hot gas atmosphere, whereas 
Baugh et~al. invoke a superwind which ejects gas that has already cooled. 
Perhaps controversially, Baugh et~al. adopt a flat IMF in starbursts, and 
a standard solar neighbourhood IMF for quiescent star formation; in 
Bower et~al., a standard IMF is assumed in all modes of star formation. 
We emphasize that, for the majority of the results presented, we have not 
adjusted any of the model parameters in order to improve the predictions 
for the fundamental plane and its projections. The one exception is where 
we exploit the modular nature of semi-analytical models to vary or switch 
off various physical ingredients of the model in order to assess their 
influence on the model predictions (Section 5.2).

The model enjoys some notable successes. We demonstrated that the model 
can match the abundance of early-type galaxies in the SDSS sample of Ber05. 
We also obtain a reasonable match to the Faber-Jackson relation between 
velocity dispersion and luminosity and its evolution with redshift, albeit 
with a shallower slope than measured by Ber05. Furthermore, we find a 
relation between velocity dispersion and age which is in excellent 
agreement with recent observations. Perhaps most impressively, the 
fundamental plane predicted by the model is in good agreement with 
that inferred for SDSS early-types by \citet{b03III}. 
The deviation from the FP relation reveals a strong correlation with 
luminosity, age, colour, stellar mass and metallicity: 
galaxies that lie above the mean fundamental plane relation are 
more luminous, younger, bluer, less massive and metal-poor. Furthermore, 
the feedback processes and clustering, as given by the pseudo-specific 
angular momentum of the bulge and the halo mass respectively, seem to 
play a role in defining the plane.

Nevertheless, despite these achievements, there are some model predictions 
which disagree with the observations. 
Formally, the slope of the predicted Faber-Jackson relation is at odds with 
that measured by Ber05, although the overlap between the model galaxies and 
observations in this projection remains impressive. 
However, perhaps the most striking discrepancy is the 
slope of the radius-luminosity relation; the model predicts a significantly 
flatter radius luminosity relation than is observed. Whereas the model 
predictions for the effective radii of faint spheroids are in good agreement 
with the data, the brightest galaxies are up to a factor of three smaller 
in the model. Our results suggest that, in the model, the brightest spheroids 
have less specific pseudo-angular momentum (i.e. 
$j_{\rm b} = r_{\rm b} \sigma$, this is a definition of convenience; see  Section~\ref{ssec:sizes}) than is the case for observed galaxies. 
This could be due to the model underpredicting the galaxy mass 
for a given luminosity. Somewhat 
surprisingly, the predicted slope of the radius-luminosity relation is in 
much better agreement with the observations if the adiabatic contraction of 
the halo is switched off (although, in this case, the model galaxies are 
uniformly too large without adjusting other parameters). The adiabatic 
contraction of the halo in response to the presence of condensed baryons has 
been tested against numerical simulations \citep[e.g.][]{jesseit, 
sellwood,choi}. 
Our prescription for computing the size of merger remnants could become 
inaccurate if there is a significant fraction of mass in the form of cold gas. 

The other significant discrepancy is the evolution with redshift of the 
zero-point of the fundamental plane. The model predicts no evolution in 
the zero-point of the fundamental plane. This is at odds with the 
evolution inferred observationally, which is consistent with the shift 
in the mass-to-light ratio expected for a passively evolving stellar 
population. This discrepancy is intriguing, as the model {\it does}  
predict a decline in the mass-to-light ratio of early-types with 
increasing redshift of the magnitude expected for passive evolution. 
The lack of evolution in the predicted fundamental plane therefore 
points to a compensating change in one of the other projections; 
the effective radii of galaxies also evolve with redshift 
in the model. This serves as a cautionary note to observational studies 
which interpret a shift in the fundamental plane in terms of a 
corresponding change in the typical mass to light ratio. 
The correlation between luminosity and the deviation from the fundamental 
plane shows that part of the evolution found by observational studies 
may in fact be due to the construction of magnitude limited samples.

In summary, the prescription outlined by Cole et~al. for computing the 
radii of discs and bulges enjoys many successes, but displays a few 
important disagreements with observations. The solution of these remaining 
problems will require enhancement of the model to compute galaxy sizes, 
guided by the results of numerical simulations of the growth of disc 
galaxies and galaxy mergers \citep[e.g.][]{okamoto, rob}.

\subsection*{ACKNOWLEDGEMENTS}
{\small
CA gratefully acknowledges a scholarship from the FCT, Portugal. 
CMB is supported by the Royal Society. This research was partly 
supported by PPARC. We thank Andrew Benson and Richard Bower 
for allowing us to use results generated with the new version of the \g~code. 
We are grateful to Mariangela Bernardi for providing the observational 
data presented in this paper. We thank Pieter van Dokkum, Alister Graham 
and the referee for valuable suggestions.

Funding for the SDSS and SDSS-II has been provided by the 
Alfred P. Sloan Foundation, the Participating Institutions, the 
National Science Foundation, the U.S. Department of Energy, the 
National Aeronautics and Space Administration, the Japanese Monbukagakusho, 
the Max Planck Society and the Higher Education Funding Council for England. 
The SDSS Web Site is  http://www.sdss.org/.

The SDSS is managed by the Astrophysical Research Consortium for the 
Participating Institutions. The Participating Institutions are the 
American Museum of Natural History, Astrophysical Institute Potsdam, 
University of Basel, Cambridge University, Case Western Reserve University, 
University of Chicago, Drexel University, Fermilab, the Institute for 
Advanced Study, the Japan Participation Group, Johns Hopkins University, 
the Joint Institute for Nuclear Astrophysics, the Kavli Institute for 
Particle Astrophysics and Cosmology, the Korean Scientist Group, the 
Chinese Academy of Sciences (LAMOST), Los Alamos National Laboratory, 
the Max-Planck-Institute for Astronomy (MPA), the Max-Planck-Institute 
for Astrophysics (MPIA), New Mexico State University, Ohio State University, 
University of Pittsburgh, University of Portsmouth, Princeton University, 
the United States Naval Observatory and the University of Washington.
}

\end{document}